\documentclass{article}
\usepackage{slashed,verbatim}
\usepackage[numbers,sort&compress]{natbib}
\usepackage{amsmath}

\newcommand{\Rmnum}[1]{\expandafter\@slowromancap\romannumeral #1@}
\makeatother
\usepackage{amssymb}
\usepackage{float}
\usepackage{comment}
\usepackage{appendix}
\usepackage{graphicx}
\usepackage{subcaption}
\usepackage{dcolumn}
\usepackage{bm}
\usepackage[colorlinks=true,linktocpage=true,
linkcolor=blue,citecolor=blue]{hyperref}
\usepackage[a4paper]{geometry}
\newcommand{\be}{\begin{eqnarray}}
\newcommand{\ee}{\end{eqnarray}}
\newcommand{\nn}{\nonumber}

\begin{document}
	\large
	\title{\bf{Thermoelectric response of a weakly magnetized thermal QCD medium }}
	\author{Debarshi Dey\footnote{ddey@ph.iitr.ac.in, debs.mvm@gmail.com}~~and~~Binoy Krishna
		Patra\footnote{binoy@ph.iitr.ac.in}\vspace{0.1in}\\
		Department of Physics,\\
		Indian Institute of Technology Roorkee, Roorkee 247667, India}
	\maketitle

\begin{abstract}
We estimate the thermoelectric response, namely, the
 Seebeck and Nernst coefficients of a hot and deconfined plasma of quarks
  and gluons, created post ultrarelativistic heavy ion collisions in the presence of a weak, homogeneous background magnetic field. 
    We employ the kinetic theory framework, wherein 
    we use the relativistic Boltzmann transport equation in the 
    relaxation time approximation. In-medium interactions are taken into account via the quasiparticle masses of the partons extracted from one loop perturbative thermal QCD. We calculate the individual and total Seebeck  
    coefficients in 2 different approaches (1-D and 2-D formulations). In the 1-D analysis, we find that a larger current quark mass has an amplifying effect on the individual Seebeck coefficient in the presence of a weak magnetic field and that the temperature sensitivities of the individual Seebeck coefficients increase with increase in the current mass of the quark species in the case of a weak magnetic field whereas the same records a decreasing trend in the presence of a strong magnetic field. The variation of individual and total Seebeck coefficients with temperature, chemical potential and background magnetic field are found to follow similar trends in both the approaches, \emph{viz.} decrease in magnitude with increasing temperature and increase in magnitude with increase in chemical potential and magnetic field. We also calculate the individual and total Nernst coefficients (in 2-D formulation) which are found to decrease with both temperature and chemical potential and increase with the magnetic field. Further, we find that the sign of the Nernst coefficient is independent of the electric charge of the charge carrier of the medium.  
\end{abstract}

\section{Introduction}
The quark-gluon plasma (QGP) is a strongly interacting state of 
matter consisting of deconfined quarks and gluons that exists under 
conditions of extremely high temperatures and/or chemical potential. 
As such, study of the properties of QGP could shed light on the evolution 
of the early universe and the structure of dense stars. At low 
temperatures ($T$) and baryon chemical potentials ($\mu_B$), QCD matter 
consists of colorless hadrons. At $\mu_B \sim 0$, with non-zero finite quark 
masses, lattice QCD results indicate that the transition from hadronic 
to quark degrees of freedom is actually a rapid analytic crossover 
rather than a true phase transition\cite{Aoki:PLB643'2006,Borsanyi:JHEP1009'2010,Borsanyi:PRD92'2015,Ding:IntJModPhysE24'2015}. At finite $\mu_B$, however, 
lattice QCD is plagued by the sign problem and hence, cannot be relied upon\cite{Hands:NuclPhysB106-107'2002,Alford:NuclPhysProcSuppl117'2003}. Compelling experimental signatures exist of the creation of QGP 
matter in Ultrarelativistic Heavy Ion collisions (URHICs) at experimental 
facilities such as the Brookhaven National Laboratory Relativistic Heavy Ion Collider (RHIC)\cite{Arsene:NuclPhysA757'2005,Adams:NuclPhysA757'2005,Adcox:NuclPhysA757'2005} and Large Hadron Collider (LHC)\cite{Carminati:JPhysG30'2004,Alessandro:JPhysG30'2006}, 
after which the created matter expands and cools, and undergoes a transition 
to a gas of interacting hadrons. The nature of the bulk evolution of QGP 
matter has been under intense investigation in the past two decades, and 
a successful description of the same has been obtained via relativistic 
hydrodynamics, which reproduced correctly the collective flow of the created matter observed in experiments. \cite{Heinz:ARNPS63'2013,Romatschke:PRL99'2007,Schenke:PRC82'2010,Niemi:PRL106'2011}. 
Using the framework of AdS/CFT correspondence, Kovtun, Son and
Starinets conjectured an extremely small, universal lower bound for the
the ratio of shear viscosity to entropy density ($\eta/s$) of $1/4\pi$ (in units with $\hbar=c=k_B=1$)\cite{Kovtun:PRL95'2005}, which makes QGP one 
of the most perfect fluids known. Hydrodynamic description of 
QGP evolution post heavy-ion collisions requires specifying 
several transport coefficients, which can be thought of as determining 
the medium's response to various perturbations. Bulk viscosity ($\zeta$) 
is expected to attain a maximum value near the QGP-Hadron gas phase 
boundary as per several lattice QCD simulations\cite{Sasaki:PRD79'2009,Sasaki:NuclPhys62'2010,Karsch:PLB217'2008}, which, in turn, affect the particle spectrum and flow coefficients\cite{Bozek:PRC81'2010,Bose:NuclPhysA931'2014}. 
The effect of thermal conductivity on the medium
has also been studied, specifically in relation to the
determination of the critical point in the QCD phase
diagram\cite{Kapusta:PRC86'2012}. 

We study the thermoelectric response of the medium which is quantified by two transport coefficients, \textit{viz.} the Seebeck coefficient and Nernst coefficient. The deconfined hot QCD medium created post heavy ion collisions can possess 
a significant temperature gradient between the central and peripheral 
regions of the collisions. Such a temperature gradient can lead to a finite 
gradient of charge carriers, resulting in an electric field- a phenomenon 
known as the Seebeck effect. 
In the presence of a temperature-gradient 
in a conducting medium, the more energetic charge
carriers in the region of higher temperature diffuse to the
region of lower temperature, leading to the generation of an
electric field. The diffusion (or equivalently, the electric current) stops 
when the created electric
field becomes strong enough to prevent further flow of
charges. The magnitude of electric field thus generated per unit temperature 
gradient in the medium is termed as the Seebeck coefficient and is evaluated 
in the limit of zero electric current\cite{Callen1960,Scheidemantel:PRB68'2003}. 
The Seebeck coefficient is a dimensionless scalar that quantifies the 
efficiency of conversion of a temperature gradient into electric field 
by a conducting medium. It is common practice to take the sign of 
the Seebeck coefficient to be positive if the direction of flow of the 
thermoelectric current is from the hotter end to the colder end. Thus, 
the sign of the Seebeck coefficient can be used to determine the sign 
of majority charge carriers in condensed matter systems, as it is 
positive for positive charge carriers and negative for negative charge 
carriers. The key parameter that gives rise to an induced current in a medium consisting of 
mobile positive and negative charge carriers, apart from a temperature gradient, is a finite chemical potential ($\mu$). This is because positive and negative charge carriers   diffuse in the same direction under the influence of a temperature gradient and as such, 
zero chemical potential (\emph{i.e.} 
equal number of particles and antiparticles) in such a medium would thus lead to 
equal and opposite electrical currents, and hence, zero net current and no Seebeck 
effect. Upcoming experimental programs such as the Facility for Antiproton 
and Ion Research (FAIR) in Germany and the  Nuclotron-based Ion Collider fAcility 
(NICA) in Russia, where low-energy heavy ion collisions are expected to create a 
baryon-rich plasma, could be the perfect environment for the aforementioned 
thermoelectric phenomenon to manifest.  Seebeck effect has been studied 
extensively in condensed matter systems such as superconductors\cite{AoArxiv,Matusiak:PRB97'2018,Hooda:EPL121'2018,Choiniere:PRX7'2017,Gaudart:PSS2185'2008}, high-temperature cuprates\cite{Seo:PRB90'2014}, organic metals\cite{Shahbazi:PRB94'2016}, etc. In the context of heavy-ion collisions, the thermoelectric response in a hadron gas has been investigated previously 
using the Hadron Resonance Gas (HRG) model\cite{Bhatt:PRD99'2018,Das:PRD102'2020}. 
Seebeck effect in a hot partonic medium has been evaluated by us in 
the absence and presence of a strong magnetic field within the relaxation time 
approximation of the relativistic Boltzmann transport equation\cite{Dey:PRD102'2020}. 

\noindent In the presence of a magnetic field, there will exist a Lorentz force on the moving charges, causing them to drift perpendicular to their original direction of motion. This transverse thermocurrent in response to a temperature gradient is called the Nernst effect. Like the Seebeck coefficient, the Nernst coefficient is also calculated at the condition of zero electric current, that is, under equilibrium conditions. The Nernst coefficient can be defined as the electric field induced in the $\hat{x}$ ($\hat{y}$) direction per unit temperature gradient in the $\hat{y}$ ($\hat{x}$) direction. In the context of heavy ion collisions, Nernst effect has been investigated in a few studies\cite{Das:PRD102'2020,kurian:PRD103'2021,Zhang:EPJC81'2021}. A comparison of the approach and results of our work with that of the other studies is also carried out in Sec. IV.B.

Large magnetic fields are created perpendicular to the plane of reaction, 
 when two charged ions collide ultrarelativistically with a finite impact 
 parameter\cite{Tuchin:AdvHEP'2013}. These fields, which depend on the center of mass energy of the collision, can be as large as $eB\sim10^{-1}
m_{\pi}^2\,(\simeq 10^{17}$ Gauss) for SPS 
energies, $eB\sim m_{\pi}^2$ for RHIC energies 
and $eB\sim 15m_{\pi}^2$ for LHC energies\cite{Sokov:IntJModPhysA24'2009}. 
Naive estimates predicted the decay of such a magnetic field to be very 
fast ($\sim$0.2 fm for RHIC energies). However, it was later pointed 
out\cite{Tuchin:PRC82'2010,Tuchin:PRC83'2011} that the finite electrical 
conductivity of the QGP medium\cite{Marty:PRC88'2013,Ding:PRD83'2011,Gupta:PLB597'2004,Amato:PRL111'2013,Aarts:PRL99'2007,Puglisi:PRD90'2014,Greif:PRD90'2014,Hattori:PRD96'2017} sustains 
the magnetic field for a much longer period of time, long enough to 
contribute significantly towards the evolution of the medium\cite{Roy:PLB750'2015,Burnier:EPJC72'2012,Mohapatra:Mod.Phys.Lett.A26'2011}. An external magnetic field in a chiral QGP medium can give rise to separation of charges, thereby breaking the $CP$ symmetry of QCD. This is called the chiral magnetic effect\cite{Kharzeev:Nucl.Phys.A803'2008,Fukushima:PRD78'2008,Kharzeev:Ann.Phys.325'2010}. Several other phenomena such as magnetic catalysis\cite{Shovkovy:LecNotesPhys871'2013}, chiral magnetic wave\cite{Kharzeev:PRD83'2011}, etc. also occur due to the presence of a background magnetic field.

The issue of the decay of the initially created magnetic field in ultrarelativistic nucleus-nucleus collisions is not a closed chapter. Several models describing the evolution of the strongly interacting quark-gluon plasma consider an infinite electric conductivity of the medium\cite{Inghirami:EPJC76'2016,Das:PRC96'2017}, which is not expected to be practically tenable. A finite, small electric conductivity of the QGP medium would cause only a small fraction of the initial magnetic field to survive by the time thermal equilibrium is achieved via interactions. This has motivated the study of several transport coefficients and other properties of the hot QCD medium in the presence of a weak magnetic field. Transport coefficients such as electric and Hall conductivities\cite{Feng:PRD96'2017,Thakur:PRD100'2019,Das:PRD101'2020,Fukushima:PRL120'2018}, shear viscosity\cite{Li:PRD97'2018} have been calculated in weak magnetic field. Thermoelectric effects in the partonic medium have been studied using the effective fugacity quasiparticle model in a weak magnetic field.\cite{kurian:PRD103'2021}. Effect of weak magnetic field on the neutral pion mass\cite{Ayala1:PRD98'2018}, quark-pion effective couplings\cite{Braghin:EPJA54'2018}, quark condensate\cite{Hofmann:PRD99'2019}, etc. have also been investigated. Further, dissociation of heavy quarkonia in the presence of weak magnetic field has also been studied recently\cite{Hasan:PRD102'2020}. Thermal transport in QCD medium has also been studied in the presence of a weak magnetic field. Thermal conductivity is related to the efficiency of heat flow or the energy
dissipation in a medium. For a hadron gas medium, thermal conductivity has been studied using the hadron gas model in the relaxation time approximation in \cite{Das:PRD100'2019}. Thermal dissipation in a deconfined medium of quarks and gluons and its interplay with charge transport processes have been investigated in \cite{Kurian:PRD102'2020}.   

In this work, we have investigated the Seebeck effect and Nernst effect in a QGP medium in 
the presence of a weak magnetic field wherein the medium interactions are encoded in the quasiparticle masses of the partons derived from one loop perturbative thermal QCD. Since temperature is the largest scale in the case of a weak background magnetic field, the quasiparticle masses have been taken to be the thermal ($B=0$) masses with magnetic field dependence appearing implicitly via the coupling constant. We make use of kinetic theory via the relativistic 
Boltzmann transport equation within the framework of relaxation time approximation
  for our study, using the electromagnetic Lorentz force field as the external 
  force term in the L.H.S. of the Boltzmann equation. It may be noted that the 
  thermalization of the matter created post heavy ion collisions is governed 
  by QCD and as such, gluons play a dominant role in the thermalization process
   since the initial density of gluons is significantly larger than that of 
   quarks or antiquarks. Magnetic field does not affect the gluons on account 
   of their electrical neutrality and hence it is a reasonable assumption to 
   consider the effect of magnetic field to be subdominant. Consequently, it can 
   be argued that the distribution function of the particles in the medium never 
   deviates significantly from equilibrium, which makes the relaxation time approximation
    of the Boltzmann transport equation, a suitable approach to calculate the Seebeck and Nernst
    coefficients (as well as other transport coefficients). Thus, in the present work, 
    we assume that the phase space and dispersion relation of the particles are 
    not affected by magnetic field via Landau quantization\cite{Feng:PRD96'2017,Thakur:PRD100'2019,Das:PRD101'2020}. The magnetic field is taken to be homogeneous and time-independent. 
    The baryon chemical potential is also considered to be homogeneous. 
    
    \noindent For our work, we have adopted two different methodologies to calculate the thermoelectric coefficients. In the first method, we have calculated only the Seebeck coefficient, considering the temperature gradient and the induced electric field to exist in the $\hat{x}$ direction only. This is made possible since the Seebeck effect is a longitudinal effect and has been done previously in multiple works\cite{Bhatt:PRD99'2018,Dey:PRD102'2020}. This has been done in order to compare the results obtained here with that of the case of strong magnetic field, already evaluated by us in an earlier work. This, thus, is the 1-dimensional evaluation of the Seebeck coefficient. Nernst effect, however, like the Hall effect, can be thought to be a transverse phenomenon since it relates the thermocurrent and temperature gradient in mutually transverse directions. This necessitates a complete 2-dimensional formulation of the problem, which at the end yields both Seebeck and Nernst coefficients simultaneously. In each approach, we have calculated first the coefficients for hypothetical media consisting of only one type of quarks, or the individual Seebeck/Nernst coefficients. Thereafter, we have calculated the coefficients for the composite medium consisting of different quark species and analysed their dependence on temperature, chemical potential and magnetic field.

The paper is organized as follows: In Sec. II, we discuss
the relativistic Boltzmann transport equation (RBTE) in the
relaxation-time approximation and set the framework for deriving 
the Seebeck and Nernst coefficients
 of the medium considering the background magnetic field to be weak. In Sec. II.A, we carry out the 1-dimensional analysis of the Seebeck coefficient and calculate both individual and total Seebeck coefficients in that framework. In Sec. II.B, we evaluate both Seebeck and Nernst coefficients using a 2-dimensional approach.
 In Sec. III, we incorporate interactions in the partonic medium via the 
 quasiparticle model by taking into account the
medium generated masses of quarks and gluons in the medium, 
arrived at by perturbative thermal QCD . In Sec. IV.A, the results of the 1-D analysis are
discussed and comparisons  with the strong magnetic field case are drawn. In sec IV.B, the results of the 2-D analysis are discussed. We finally conclude in section V.



\section{Seebeck and Nernst coefficients from the relaxation time approximation of the Boltzmann equation}
In this section, we develop a general framework to study the thermoelectric response of a quark gluon plasma medium quantified by the Seebeck and Nernst coefficients, in the presence of a weak, homogeneous background magnetic field using the Boltzmann transport equation in the relaxation time approximation. 

Each parton in the plasma is associated with a one-particle distribution function, $f(x,p)\equiv f(\vec{r},\vec{p},t)$ which is a Lorentz invariant density in phase space, so that $f(\vec{r},\vec{p},t)\,d^3r\,d^3p$ gives the number of partons in the spatial volume element $d^3r$ about $\vec{r}$ and with momenta in a range $d^3p$ about $\vec{p}$. The evolution of this distribution function towards equilibrium via collisions is described by the Boltzmann transport equation
\begin{align}
\frac{d f}{dt}&=\frac{\partial f}{\partial t} +\frac{\vec{p}}{m} 
\cdot \nabla f + {\vec{F}} \cdot \frac{\partial f}{\partial \vec{p}}
\nonumber\\
&= {\left( \frac{\partial f}{\partial t} \right)}_{\rm coll},
\label{bte}
\end{align}
where ${\vec{F}}$ is the force field acting on the particles
in the medium. If the collision term is zero
then the particles do not collide, and individual collisions get
replaced by long-range aggregated (Coulomb) interactions, and the equation is then referred to
as the collisionless Boltzmann equation or Vlasov equation. Particles arrive at and leave from the phase space volume element under consideration, as a result of collisions. With both these processes is associated the matrix element $M_{pp^{\prime}}$ such that $\frac{M_{\vec{p}\vec{p}^{\prime}}dt\,dp^{\prime}}{{(2\pi)}^3}$ is the probability that a parton with momentum $\vec{p}$ is scattered into an infinitesimal volume element $dp^{\prime}$ about $p^{\prime}$ in the time-interval $dt$. The probability per unit time that a parton
with momentum $\vec{p}$ suffers some collision is obtained by summing over final state momenta $\vec{p^{\prime}}$:
\begin{equation}
\frac{1}{\tau(\vec{p})}=\int\frac{d \vec{p^\prime}}{(2\pi)^3}~M_{\vec{p}
	\vec{p^\prime}} 
[1-f(\vec{p^\prime})],\label{rti}
\end{equation}
The factor $(1-f (\vec{p^\prime}))$ denotes the reduction in available final states imposed by the exclusion principle. Therefore, the total
number of quarks with momenta in the neighbourhood of $\vec{p}$ that suffer collision in time $dt$, thereby moving out of the concerned phase space volume element, is
\begin{equation}
\frac{dt}{\tau(\vec{p})}f(\vec{p})\frac{d \vec{p}}{(2\pi)^3}. \label{4*}
\end{equation}
By defining appropriately the factor ${\frac{df(\vec{p})}{dt}|}_{\rm out}$, the above quantity can be re-expressed as
\begin{equation}
-{\frac{df(\vec{p})}{dt}\Big|}_{\text{out}}\frac{d\vec{p}}{(2\pi)^3}dt.
\label{3*}
\end{equation}
Equating with Eq.\eqref{4*} yields:
\begin{align}
\frac{df(\vec{p})}{dt}\Big|_{\text{out}} &=-\frac{f(\vec{p})}{\tau 
	(\vec{p})}
\nonumber\\
&=-f(\vec{p})\int\frac{d\vec{p^\prime}}{(2\pi)^3} 
M_{\vec{p} \vec{p^\prime}}[1-f(\vec{p^\prime})].\label{5*}
\end{align}  
Similarly, the total number of partons that enter the phase space volume element (acquire momenta in the range $dp$ about $\vec{p}$) as a result of collisions in time interval dt is given by:
\begin{equation}
\frac{df(\vec{p})}{dt}\Big|_{\text{in}}=[1-f(\vec{p})]\int\frac{d 
	\vec{p^\prime}}{(2\pi)^3}
M_{\vec{p^\prime}\vec{p}}[f(p^{\prime})].\label{6*}
\end{equation}
Finally, the collision integral yields a general form :
\begin{align}
\left(\frac{df(\vec{p})}{dt}\right)_{\text{coll}} &= \frac{df(\vec{p})}{dt}
\Big|_{\text{in}}
+\frac{df(\vec{p})}{dt}\Big|_{\text{out}}\nonumber\\
&=-\int\frac{d p^{\prime}}{(2\pi)^3}
\{M_{\vec{p}\vec{p^\prime}}f(\vec{p})[1-f(\vec{p^\prime})]-
M_{\vec{p^\prime}\vec{p}}f(\vec{p^\prime}) 
[1-f(\vec{p})]\},\label{7*}
\end{align}
which makes the Boltzmann equation \eqref{bte} a non-linear integro-differential 
equation and is therefore difficult to solve in this generic form. To make the Boltzmann equation tractable, we resort to what is called the relaxation-time approximation (RTA). The RTA helps to linearize the Boltzmann equation by virtue of a set of assumptions: 

\noindent i) The distribution of partons emerging from collisions 
at any time does not 
depend on the structure of the non equilibrium distribution function 
$f(\vec{r},\vec{p},t)$ just prior to collisions.

 \noindent ii) If the partons in a 
region about $\vec{r}$ have the equilibrium distribution 
appropriate to a local temperature $T(\vec{r})$,
\[f(\vec{p})=f_0(\vec{p})=\frac{1}{\text{exp}
	\left(\frac{\epsilon-\mu}{T(\vec{r})}\right)\pm 1},\]
then, collisions will not alter the form of the distribution function.	Consequently, the probability per unit time for a collision $1/\tau(p)$, becomes a specified function of $\vec{p}$ without any dependence on $f(\vec{r},\vec{p},t)$, unlike Eq.\eqref{rti}. Thus, the {\em out}-term in the collision
integral gets simplified as
\begin{equation}
\frac{df(\vec{p})}{dt}\Big|_{\text{out}} =-\frac{f(\vec{p})}{\tau(\vec{p})}.
\end{equation}
The {\em in}-term that signifies the number of partons entering the concerned phase space volume element as a result of collisions, now involve the equilibrium distribution function and is given by:
 \begin{equation}
\frac{df(\vec{p})}{dt}\Big|_{\text{in}}=\frac{f_0(\vec{p})}{\tau(\vec{p})}.
\end{equation}
Thus,  the collision integral (and hence the transport equation) is linearized via RTA in the form
\begin{align}
C[f] &= -\frac{\left(f(\vec{p})-f_0(\vec{p})\right)}{\tau(\vec{p})},\nonumber\\
&= -\frac{\delta f}{\tau (\vec{p})},\label{10*}
\end{align} 
 A slight deviation of the system from equilibrium is taken into account by allowing for a relatively small $\delta f$ in Eq.\eqref{10*}, compared to the local equilibrium distribution function $f_0$, $i.e.\quad f=f_0+\delta f$, with $\delta f\ll f_0$. $\tau(p)$ is the relaxation-time of the medium. Since a deconfined medium of quarks and gluons is a relativistic system, it makes sense to work with the relativistic Boltzmann transport equation which reads for the $i$-th parton: 
\begin{equation}
p^{\mu}\frac{\partial f_i(x,p)}{\partial 
x^{\mu}}+q_iF^{\mu \nu}p_{\nu}\
\frac{\partial f_i(x,p)}{\partial p^{\rho}}
=C[f_i(x,p)],\label{1}
\end{equation}
where, $f_i(x,p)$ and $q_i$ are the distribution function and electric charge,  respectively, of the $i$th quark flavour, $F^{\mu\nu}$ is the
electromagnetic field strength tensor and $C[f_i(x,p)]$ is the collision term already discussed earlier.  
To study the Seebeck effect, we need consider only the quarks, as gluons do not contribute to the electric current. Thus, the equilibrium one-particle distribution function for a plasma moving with a macroscopic four-velocity $u^{\nu}$ is the Fermi-Dirac distribution given by:
\be f_i^0(\vec{r},\vec{p})\equiv f_i^0=\frac{1}{\mbox{exp}(\frac{u_{\nu}p^{\nu}-\mu_i}{T})-1},
\ee
where, $\mu$ refers to the quark chemical potential. In the local rest frame of the plasma, $u^{\nu}=(1,0,0,0)$ and the distribution function reduces to:
\be f_i^0=\frac{1}{\mbox{exp}(\frac{\epsilon_i-\mu_i}{T})-1},\label{df}
\ee
with $\epsilon(\vec{p})=\sqrt{\vec{p}^2+m^2}$ and $\beta(\vec{r})=1/T(\vec{r})$. The collision term under RTA is simplified to
\be
C[f_i(x,p)]\simeq -\frac{p^{\mu}u_{\mu}}
{\tau_i}\,\delta f_i.\label{ct}
\ee
Eq.\eqref{1} can then be written as
\be
p^{\mu}\frac{\partial f_i(x,p)}{\partial 
x^{\mu}}+ F^{\prime\mu}\frac{\partial f_i(x,p)}{\partial p^{\mu}}=-\frac{p^{\mu}u_{\mu}}
{\tau_i}\,\delta f_i.\label{2}
\ee
Here, $F^{\prime\mu}=(p^0\vec{v}.\vec{F},p^0\vec{F})$ is a 4-vector that can be thought of as the relativistic counterpart of the classical electromagnetic force with  $\vec{F}=q(\vec{E}+\vec{v}\times \vec{B})$ being the background classical electromagnetic force field. By using $F^{0i}=-E^i$ and $2F_{ij}=\epsilon_{ijk}B^k$, we can show:
\be F^{\prime\mu}=qF^{\mu\nu}p_{\nu},\label{3}
\ee
where, $\epsilon_{ijk}$ is the completely antisymmetric Levi-Civita tensor.
Writing RBTE in 3-notation, we have (dropping the particle label $i$):
\be \left(p^0\frac{\partial}{\partial x^0}+p^j\frac{\partial}{\partial x^j}+F^0\frac{\partial}{\partial p^0}+F^j\frac{\partial}{\partial p^j}\right)f=-\frac{p^0}{\tau}(f-f_0).\label{4}\ee
$i.e.$
\be \frac{\partial f}{\partial t}+\vec{v}.\frac{\partial f}{\partial \vec{r}}+\frac{\vec{F}.\vec{p}}{p^0}\frac{\partial f}{\partial p^0}+\vec{F}.\frac{\partial f}{\partial \vec{p}}=-\frac{(f-f_0)}{\tau}.\label{5}
\ee
Under steady state assumption, $\frac{\partial f}{\partial t}=0$. Thus, we get:
\be \left(\vec{v}.\frac{\partial }{\partial \vec{r}}+\frac{\vec{F}.\vec{p}}{p^0}\frac{\partial }{\partial p^0}+\vec{F}.\frac{\partial }{\partial \vec{p}}\right)f=-\frac{(f-f_0)}{\tau}.\label{6}
\ee
Considering $p^0$ to be an independent variable, we make use of the chain rule
\begin{equation}
	\frac{\partial}{\partial \vec{p}}\rightarrow \frac{\partial p^0}{\partial \vec{p}}\frac{\partial}{\partial p^0}+\frac{\partial}{\partial \vec{p}}=\frac{\vec{p}}{p^0}\frac{\partial}{\partial p^0}+\frac{\partial}{\partial \vec{p}}.\label{cr}
\end{equation}
Thus, Eq.\eqref{6} becomes
\begin{equation}
\vec{v}.\frac{\partial f}{\partial \vec{r}}+\vec{F}.\frac{\partial f}{\partial \vec{p}}=-\frac{(f-f_0)}{\tau},\label{8}
\end{equation}
with $\vec{F}=q(\vec{E}+\vec{v}\times \vec{B})$.

In what follows, we evaluate the Seebeck coefficient in a weak magnetic field, first in a one-dimensional formulation and carry out a comparison with the Seebeck coefficient in the presence of a strong magnetic field, evaluated in our previous work. A one-dimensional treatment is possible since the Seebeck effect is a longitudinal effect\cite{Bhatt:PRD99'2018}. Thereafter, we carry out a two-dimensional evaluation of the thermoelectric response in a weak magnetic field which yields simultaneously the Seebeck and Nernst coefficients.
 \subsection{One-dimensional formulation: Seebeck coefficient} 
 let us consider $\vec{E}=E\, \hat{x}$, $\vec{B}=B\,\hat{z}$. Then, eq.\eqref{8} becomes:
\be  f-qB\tau\left(v_x\frac{\partial f}{\partial p_y}-v_y\frac{\partial f}{\partial p_x}\right)=f_0-qE\tau \frac{\partial f_0}{\partial p_x}-\tau \vec{v}.\frac{\partial f}{\partial \vec{r}}.\label{11}\ee
To solve eq.\eqref{11}, we employ the following ansatz\cite{Feng:PRD96'2017}:
\be f=f_0-\tau qE\frac{\partial f_0}{\partial p_x}-\vec{\chi}.\frac{\partial f_0}{\partial \vec{p}},\label{12}\ee
$f_0$ being the equilibrium distribution function defined in Eq.\eqref{df}.

Thus, Eq.\eqref{11} becomes:
\be -\vec{\chi}.\frac{\partial f_0}{\partial \vec{p}}-qB\tau\left(v_x\frac{\partial f}{\partial p_y}-v_y\frac{\partial f}{\partial p_x}\right)=-\tau \vec{v}.\frac{\partial f}{\partial \vec{r}}.\label{13}\ee


Simplifying the terms in the parenthesis using the ansatz given in Eq.\eqref{12}, we get,
\begin{equation}
v_x\frac{\partial f}{\partial p_y}
=v_x\left\{\frac{\partial f_0}{\partial p_y}-\tau qE\frac{\partial ^2f_0}{\partial p_y\partial p_x}-\chi_x\frac{\partial ^2f_0}{\partial p_y\partial p_x}-\chi_y\frac{\partial ^2f_0}{\partial p_y^2}-\chi_z\frac{\partial ^2f_0}{\partial p_y\partial p_z}\right\}.\label{14}
\end{equation}

\begin{equation}
v_y\frac{\partial f}{\partial p_x}=v_y\left\{\frac{\partial f_0}{\partial p_x}-\tau qE\frac{\partial ^2f_0}{\partial p_x^2}-\chi_y\frac{\partial ^2f_0}{\partial p_x\partial p_y}-\chi_x\frac{\partial ^2f_0}{\partial p_x^2}-\chi_z\frac{\partial ^2f_0}{\partial p_x\partial p_z}\right\}.\label{15}
\end{equation}
Using Eq.\eqref{df},
\be \frac{\partial f_0}{\partial p_x}=\beta f_0(f_0-1)\frac{p_x}{\epsilon}.\label{17}\ee

We have,
\begin{align}
\frac{\partial^2f_0}{\partial p_y\partial p_x}&=\frac{\beta f_0p_xp_y}{\epsilon^2}\left(\beta+\frac{1}{\epsilon}\right),\label{21}\\[0.5em]
\frac{\partial^2f_0}{\partial p_y\partial p_z}&=\frac{\beta f_0p_yp_z}{\epsilon^2}\left(\beta+\frac{1}{\epsilon}\right),\label{22}\\[0.5em]
\frac{\partial^2f_0}{\partial p_y^2}&=-\frac{\beta f_0}{\epsilon}\left(1-\frac{p_y^2}{\epsilon^2}-\frac{\beta p_y^2}{\epsilon}\right),\label{23}\\[0.5em]
\frac{\partial^2f_0}{\partial p_x^2}&=-\frac{\beta f_0}{\epsilon}\left(1-\frac{p_x^2}{\epsilon^2}-\frac{\beta p_x^2}{\epsilon}\right),\label{24}
\end{align}
where we have neglected $f_0^2$ at high $T$.

\noindent Making use of equations \eqref{21}, \eqref{22}, \eqref{23}, \eqref{24} in expressions \eqref{14} and \eqref{15}, and retaining terms with only linear power of velocity\cite{Feng:PRD96'2017}, we get:
\begin{equation*} 
v_x\frac{\partial f}{\partial p_y}-v_y\frac{\partial f}{\partial p_x}=\frac{\partial f_0}{\partial \epsilon}\frac{1}{\epsilon}\left(\chi_xv_y-\chi_yv_x+\tau qEv_y\right),\label{27}
\end{equation*}
where we have used $\frac{\partial f_0}{\partial \epsilon}=-\beta(f_0-f_0^2)\simeq -\beta f_0$.

Thus, the L.H.S. of Eq.\eqref{13} becomes:
\be -\vec{\chi}.\frac{\partial f_0}{\partial \vec{p}}-q^2B\tau^2E\,\frac{\partial f_0}{\partial \epsilon}\frac{v_y}{\epsilon}+qB\tau\,\frac{\partial f_0}{\partial \epsilon}\frac{1}{\epsilon}\left(v_x\chi_y-v_y\chi_x\right).\label{28}
\ee
We now simplify the R.H.S. of Eq.\eqref{13}, which, on using the ansatz becomes:
\begin{equation}
-\tau \vec{v}\cdot\left[\frac{\partial f_0}{\partial \vec{r}}-\tau qE\frac{\partial}{\partial \vec{r}}\left(\frac{\partial f_0}{\partial p_x}\right)-\chi_x\frac{\partial}{\partial \vec{r}}\left(\frac{\partial f_0}{\partial p_x}\right)-\chi_y \frac{\partial}{\partial \vec{r}}\left(\frac{\partial f_0}{\partial p_y}\right)-\chi_z \frac{\partial}{\partial \vec{r}}\left(\frac{\partial f_0}{\partial p_z}\right) \right].\label{n1}
\end{equation}
We consider only the temperature gradient in the $x$ direction, $\frac{\partial T}{\partial x}$. Using
$$\frac{\partial f_0}{\partial \vec{p}}=\frac{\partial f_0}{\partial \epsilon}\frac{\partial \epsilon}{\partial \vec{p}}=\frac{\partial f_0}{\partial \epsilon}\frac{\vec{p}}{\epsilon},$$
and
$$\frac{\partial f_0}{\partial \vec{r}}=-\frac{\partial f_0}{\partial \epsilon}\left(\frac{\epsilon-\mu}{T}\right)\vec{\nabla}_{\vec{r}}\,T,$$
and on neglecting higher order velocity terms, the R.H.S. simplifies to:
\begin{equation}
\tau v_x \left(\frac{\partial f_0}{\partial \epsilon}\right)\left(\frac{\epsilon-\mu}{T}\right)\frac{\partial T}{\partial x}.\label{n3}
\end{equation}

Equating the L.H.S. and R.H.S. from Eq.\eqref{28} and Eq.\eqref{n3} respectively and dividing throughout by $\tau$, we finally obtain
\be  v_x\left[\frac{-\chi_x}{\tau}+\frac{qB\chi_y}{\epsilon}-\left(\frac{\epsilon-\mu}{T}\right)\frac{\partial T}{\partial x}\right]+v_y\left[\frac{-\chi_y}{\tau}-\frac{qB\chi_x}{\epsilon}-\frac{q^2B\tau E}{\epsilon}\right]+v_z\left[\frac{-\chi_z}{\tau}\right]=0.\label{30}\ee
Calling $\frac{qB}{\epsilon}=\omega_c$ as the cyclotron frequency, we compare the coefficients of $v_x$, $v_y$ and $v_z$ on both sides of eq.\eqref{30}, to get:
\begin{align}
\frac{\chi_x}{\tau}-\omega_c\chi_y+\left(\frac{\epsilon-\mu}{T}\right)\frac{\partial T}{\partial x}=&0.\label{31a}\\[0.5em]
\frac{\chi_y}{\tau}+\omega_c\chi_x+\omega_cq\tau E=&0.\label{31b}\\[0.4em]
\frac{\chi_z}{\tau}=&0.\label{31c}
\end{align}

Solving the equations above for $\chi_x$, $\chi_y$ and $\chi_z$, we get:
\begin{align}
\chi_x=&\frac{-\omega_c^2\tau^3}{1+\omega_c^2\tau^2}qE-\frac{\tau}{1+\omega_c^2\tau^2}\left(\frac{\epsilon-\mu}{T}\right)\frac{\partial T}{\partial x}.\label{32a}\\[0.5em]
\chi_y=&\frac{-\omega_c\tau^2}{1+\omega_c^2\tau^2}qE+\frac{\omega_c\tau^2}{1+\omega_c^2\tau^2}\left(\frac{\epsilon-\mu}{T}\right)\frac{\partial T}{\partial x}.\label{32b}\\[0.5em]
\chi_z=&0.\label{32c}
\end{align}
Substituting Eq.\eqref{32a}, Eq.\eqref{32b} and Eq,\eqref{32c} in the ansatz [Eq.\eqref{12}], we obtain:
\begin{align}
\delta f&=\left[-\tau qv_x+\frac{\omega_c^2\tau^3}{1+\omega_c^2\tau^2}qv_x+\frac{\omega_c\tau^2}{1+\omega_c^2\tau^2}qv_y\right]\frac{\partial f_0}{\partial \epsilon}E\nonumber\\
&-\left[\frac{\omega_c\tau^2}{1+\omega_c^2\tau^2}\left(\frac{\epsilon-\mu}{T}\right)v_y-\frac{\tau}{1+\omega_c^2\tau^2}\left(\frac{\epsilon-\mu}{T}\right)v_x\right]\frac{\partial f_0}{\partial \epsilon}\frac{\partial T}{\partial x}.\label{35}
\end{align}
For antiparticles, 
\begin{align}
\overline{\delta f}&=\left[-\tau \bar{q}v_x+\frac{\omega_c^2\tau^3}{1+\omega_c^2\tau^2}\bar{q}v_x+\frac{\bar{\omega_c}\tau^2}{1+\omega_c^2\tau^2}\bar{q}v_y\right]\frac{\partial \bar{f_0}}{\partial \epsilon}E\nonumber\\
&-\left[\frac{\bar{\omega_c}\tau^2}{1+\omega_c^2\tau^2}\left(\frac{\epsilon-\mu}{T}\right)v_y-\frac{\tau}{1+\omega_c^2\tau^2}\left(\frac{\epsilon-\mu}{T}\right)v_x\right]\frac{\partial \bar{f_0}}{\partial \epsilon}\frac{\partial T}{\partial x}.\label{36}
\end{align}
The induced four
current is then obtained from the relation
\be J^{\mu}=qg\int \frac{d^3\mbox{p}}{(2\pi)^3\epsilon}p^{\mu}\left[\delta f-\overline{\delta f}\right],\label{38}
\ee
where $q$ and $g$ are respectively the electric charge and degeneracy factor of the quark species. Substituting in Eq.\eqref{38} the relevant terms from Eq.\eqref{35} and Eq.\eqref{36}, we thus obtain the spatial part of the induced four
current, i.e., the induced current density
\begin{multline} 
J_x=\frac{qg\times 4\pi}{8\pi^3}\int \frac{\mbox{dp}}{\epsilon^2}p^4q\frac{\tau \beta}{(1+\omega_c^2\tau^2)}\left\{f_0(1-f_0)+\bar{f_0}(1-\bar{f_0})\right\}E\\+\frac{qg\times 4\pi}{8\pi^3}\int \frac{d\mbox{p}}{\epsilon^2}p^4\frac{\tau \beta}{T(1+\omega_c^2\tau^2)}\left\{(\epsilon+\mu)\bar{f_0}(1-\bar{f_0})-(\epsilon-\mu)f_0(1-f_0)\right\}\frac{\partial T}{\partial x}.\label{41}
\end{multline}
Setting $J_x=0$, we obtain:
\begin{align}
E&=\frac{1}{qT}\frac{\int \mbox{dp}\frac{p^4\tau}{\epsilon^2(1+\omega_c^2\tau^2)}\left\{(\epsilon-\mu)f_0(1-f_0)-(\epsilon+\mu)\bar{f_0}(1-\bar{f_0})\right\}}{\int \mbox{dp} \frac{p^4\tau}{\epsilon^2(1+\omega_c^2\tau^2)}\left\{f_0(1-f_0)+\bar{f_0}(1-\bar{f_0})\right\}}\frac{\partial T}{\partial x}\nn\\[0.8em]
&=\frac{1}{qT}\frac{I_2}{I_1}\frac{\partial T}{\partial x}.
\end{align}
where,
\begin{align}
I_1&=\int \mbox{dp} \frac{p^4\tau}{\epsilon^2(1+\omega_c^2\tau^2)}\left\{f_0(1-f_0)+\bar{f_0}(1-\bar{f_0})\right\}\nn\\
I_2&=\int \mbox{dp}\frac{p^4\tau}{\epsilon^2(1+\omega_c^2\tau^2)}\left\{(\epsilon-\mu)f_0(1-f_0)-(\epsilon+\mu)\bar{f_0}(1-\bar{f_0})\right\}.
\end{align}
 Equating with $(\vec{E})_x=S\Big(\vec{\nabla}_{\vec{r}}\,T(r)\Big)_x$, we obtain:
 \begin{equation}
S=\frac{1}{qT}\frac{I_2}{I_1}.\label{43}
 \end{equation}
 

After having calculated the Seebeck coefficient 
for a thermal medium consisting of a single 
species, we move on to the more realistic 
case of a multi-component system, which in 
our case corresponds to multiple flavours of 
quarks in the QGP. However, gluons being electrically 
neutral, do not contribute to the thermoelectric 
current, therefore, the total electric current in the medium is 
the vector sum of currents due to individual species:
\begin{align}
J_x&=(J_x)_1+(J_x)_2+(J_x)_3+ \cdots \nonumber\\
&=\left(\frac{q_1^2g_1}{2T\pi^2}(I_1)_1+\frac{q_2^2g_2}{2T\pi^2}(I_1)_2+...
\right)E-\bigg(\frac{q_1g_1}
{2T^2\pi^2}(I_2)_1+\frac{q_2g_2}{2T^2
	\pi^2}(I_2)_2+...\bigg)\frac{\partial T}{\partial x}.\label{44}
\end{align}
Setting the total current, $J_x=0$ as earlier, we get the 
induced electric field,
\begin{equation}
E=\frac{\sum_{i}\frac{q_ig_i
		(I_2)_i}{T}}{\sum_{i}q_i^2g_i
	(I_1)_i}\frac{\partial T}{\partial x}.
\label{45}
\end{equation}
All quarks have the same degeneracy factor. 
Hence, the total Seebeck coefficient for the multi-component 
system can be rewritten as
\begin{equation}
S=\frac{\sum_{i} S_i\,q_i^2(I_1)_i}{\sum_
	{i}q_i^2(I_1)_i},\label{47}
\end{equation}
which could be viewed as a weighted
average of the Seebeck coefficients of 
individual species ($S_i$) present in the medium.

\subsection{Two-dimensional formulation: Seebeck and Nernst coefficients}
A conducting medium subjected to mutually perpendicular magnetic field and temperature gradient develops a thermocurrent perpendicular to both the magnetic field and temperature gradient. This phenomenon is called the Nernst effect. While the Seebeck coefficient determined from the `open circuit' condition relates the electric field component in a particular direction to the temperature gradient component in the same direction, the Nernst coefficient can be thought of as a Hall type thermoelectric coefficient that relates the electric field and the temperature gradient in mutually transverse directions. Thus, evaluating the Nernst coefficient requires a 2-dimensional formulation of the problem. Here, we consider the electric field and temperature gradient to exist in the $x$-$y$ plane with the magnetic field pointing exclusively in the $z$ direction. Also, we consider a two flavour quark gluon plasma medium with $u$ and $d$ quarks (and their antiquarks). We first evaluate the Seebeck and Nernst coefficients for a QGP medium composed of a single quark species.

With $\vec{E}=E_x\,\hat{x}+E_y\,\hat{y}$, the Boltzmann equation [Eq.\eqref{8}] reads:
\begin{equation}
f-qB\tau\left(v_x\frac{\partial f}{\partial p_y}-v_y\frac{\partial f}{\partial p_x}\right)=f_0-\tau \vec{v}\cdot \frac{\partial f}{\partial \vec{r}}-\tau q\vec{E}\cdot\frac{\partial f}{\partial \vec{p}}\,,\label{48}
\end{equation}
where, $f_0$ is the equilibrium quark distribution function given by Eq.\eqref{df} and $f$ is the total distribution function satisfying $f=f_0+\delta f$. We modify the ansatz in Eq.\eqref{12} to include $E_y$ :
\begin{equation}
f=f_0+\delta f=f_0-\tau q\vec{E}\cdot\frac{\partial f_0}{\partial \vec{p}}-\vec{\chi}.\frac{\partial f_0}{\partial \vec{p}}.\label{49}
\end{equation}
Using Eq.\eqref{49}, Eq.\eqref{48} becomes:
\begin{equation}
\vec{\chi}\cdot \frac{\partial f_0}{\partial \vec{p}}-qB\tau\left(v_y\frac{\partial f}{\partial p_x}-v_x\frac{\partial f}{\partial p_y}\right)=\tau \vec{v}\cdot \frac{\partial f_0}{\partial \vec{r}}.\label{50}
\end{equation}
The terms in the parenthesis, after using the ansatz and retaining only linear velocity terms, simplify to
\begin{equation}
v_y\frac{\partial f}{\partial p_x}-v_x\frac{\partial f}{\partial p_y}=v_y\chi_x+v_y\tau qE_x-v_x\chi_y-v_x\tau qE_y.
\end{equation}
This finally leads to
\begin{equation}
v_x\left[\frac{\chi_x}{\tau}-\omega_c\tau qE_y-\omega_c\chi_y+\frac{\epsilon-\mu}{T}\frac{\partial T}{\partial x}\right]+v_y\left[\frac{\chi_y}{\tau}+\omega_c\tau qE_x+\omega_c\chi_x+\frac{\epsilon-\mu}{T}\frac{\partial T}{\partial y}\right]=0,\label{51}
\end{equation}
where, $\omega_c=qB/\epsilon$ is the cyclotron frequency. Equating coefficients of $v_x$ and $v_y$, we get
\begin{align}
\frac{\chi_x}{\tau}-\omega_c\tau qE_y-\omega_c\chi_y+\frac{\epsilon-\mu}{T}\frac{\partial T}{\partial x}&=0.\\[0.4em]
\frac{\chi_y}{\tau}+\omega_c\tau qE_x+\omega_c\chi_x+\frac{\epsilon-\mu}{T}\frac{\partial T}{\partial y}&=0.
\end{align}
Solving for $\chi_x$ and $\chi_y$ yields:
\begin{align}
\chi_x=&\frac{-\omega_c^2\tau^3}{1+\omega_c^2\tau^2}qE_x-\frac{\tau}{1+\omega_c^2\tau^2}\left(\frac{\epsilon-\mu}{T}\right)\frac{\partial T}{\partial x}+\frac{\omega_c\tau^2}{1+\omega_c^2\tau^2}qE_y-\left(\frac{\epsilon-\mu}{T}\right)\frac{\omega_c\tau^2}{1+\omega_c^2\tau^2}\frac{\partial T}{\partial y}.\\[0.5em]
\chi_y=&\frac{-\omega_c\tau^2}{1+\omega_c^2\tau^2}qE_x+\frac{\omega_c\tau^2}{1+\omega_c^2\tau^2}\left(\frac{\epsilon-\mu}{T}\right)\frac{\partial T}{\partial x}-\frac{\omega_c^2\tau^3}{1+\omega_c^2\tau^2}qE_y-\frac{\tau}{1+\omega_c^2\tau^2}\left(\frac{\epsilon-\mu}{T}\right)\frac{\partial T}{\partial y}.
\end{align}
Substituting in Eq.\eqref{49}, we obtain:
\begin{align}
\delta f&= \frac{\partial f_o}{\partial \epsilon}\left[-\tau qv_x+\frac{\omega_c^2\tau^3}{1+\omega_c^2\tau^2}qv_x+\frac{\omega_c\tau^2}{1+\omega_c^2\tau^2}qv_y\right]E_x+\frac{\partial f_o}{\partial \epsilon}\Big[-\tau qv_y+\frac{\omega_c^2\tau^3}{1+\omega_c^2\tau^2}qv_y\nonumber\\[0.5em]&-\frac{\omega_c\tau^2}{1+\omega_c^2\tau^2}qv_x\Big]E_y+\frac{\partial f_o}{\partial \epsilon}\left[\frac{\tau}{1+\omega_c^2\tau^2}\left(\frac{\epsilon-\mu}{T}\right)v_x-\frac{\omega_c\tau^2}{1+\omega_c^2\tau^2}\left(\frac{\epsilon-\mu}{T}\right)v_y\right]\frac{\partial T}{\partial x}\nonumber\\[0.5em]&+\frac{\partial f_o}{\partial \epsilon}\left[\frac{\tau}{1+\omega_c^2\tau^2}\left(\frac{\epsilon-\mu}{T}\right)v_y+\frac{\omega_c\tau^2}{1+\omega_c^2\tau^2}\left(\frac{\epsilon-\mu}{T}\right)v_x\right]\frac{\partial T}{\partial y}.\label{52}
\end{align}
$\overline{\delta f}$ is obtained by replacing $q$ by $-q$ in Eq.\eqref{52}. The induced 4-current as earlier is given by:
\be
J^{\mu}=qg\int \frac{d^3\mbox{p}}{(2\pi)^3\epsilon}p^{\mu}\left[\delta f-\overline{\delta f}\right].\label{53}
\ee
Substituting the expressions for $\delta f$ and $\overline{\delta f}$ above, we obtain:
\begin{align}
J_x&=\frac{qg}{6\pi^2}\left[(q\beta I_1)E_x+(q\beta I_2)E_y+(\beta^2I_3)\frac{\partial T}{\partial x}+(\beta^2I_4)\frac{\partial T}{\partial y}\right],\\[0.5em]
J_y&=\frac{qg}{6\pi^2}\left[(q\beta I_1)E_y+(-q\beta I_2)E_x+(\beta^2I_3)\frac{\partial T}{\partial y}+(-\beta^2I_4)\frac{\partial T}{\partial x}\right],
\end{align}
where,
\begin{align}
I_1&=\int \mbox{dp}\,p^4 \frac{\tau}{\epsilon^2(1+\omega_c^2\tau^2)}\left\{f_0(1-f_0)+\bar{f_0}(1-\bar{f_0})\right\}\nn\\[0.4em]
I_2&=\int \mbox{dp}\,p^4 \frac{\omega_c\tau^2}{\epsilon^2(1+\omega_c^2\tau^2)}\left\{f_0(1-f_0)-\bar{f_0}(1-\bar{f_0})\right\}\nn\\[0.4em]
I_3&=\int \mbox{dp}\,p^4\frac{\tau}{\epsilon^2(1+\omega_c^2\tau^2)}\left\{(\epsilon+\mu)\bar{f_0}(1-\bar{f_0})-(\epsilon-\mu)f_0(1-f_0)\right\}\nn\\[0.4em]
I_4&=\int \mbox{dp}\,p^4\frac{\omega_c\tau^2}{\epsilon^2(1+\omega_c^2\tau^2)}\left\{-(\epsilon+\mu)\bar{f_0}(1-\bar{f_0})-(\epsilon-\mu)f_0(1-f_0)\right\}\nn
\end{align}  
In equilibrium, we have, $J_x=0=J_y$. This leads to
\begin{align}
C_1E_x+C_2E_y+C_3\frac{\partial T}{\partial x}+C_4\frac{\partial T}{\partial y}&=0\label{54},\\[0.5em]
-C_2E_x+C_1E_y-C_4\frac{\partial T}{\partial x}+C_3\frac{\partial T}{\partial y}&=0,\label{55}
\end{align}
where, $C_1=qI_1$, $C_2=qI_2$, $C_3=\beta I_3$ and $C_4=\beta I_4$.

\noindent The electric field components are related to the components of the temperature gradients via the Seebeck and Nernst coefficients via a matrix equation
\begin{eqnarray}
\begin{pmatrix}
E_x\\
E_y
\end{pmatrix}=\begin{pmatrix}
S & N|\vec{B}|\\
-N|\vec{B}| & S
\end{pmatrix}\begin{pmatrix}
\frac{\partial T}{\partial x}\\
\frac{\partial T}{\partial y}
\end{pmatrix}.
\end{eqnarray}
Here, $S$ and $N$ refer to the Seebeck and Nernst coefficients respectively. The relative minus sign among the Nernst coefficients is necessitated by the Onsager reciprocity theorem\cite{Callen1960}. Using Eq.\eqref{54} and Eq.\eqref{55}, we finally obtain:
\begin{align}
E_x=\left[-\frac{C_1C_3+C_2C_4}{C_1^2+C_2^2}\right]\frac{\partial T}{\partial x}+\left[\frac{C_2C_3-C_1C_4}{C_1^2+C_2^2}\right]\frac{\partial T}{\partial y}\label{56}\\[0.5em]
E_y=\left[-\frac{C_1C_3+C_2C_4}{C_1^2+C_2^2}\right]\frac{\partial T}{\partial y}-\left[\frac{C_2C_3-C_1C_4}{C_1^2+C_2^2}\right]\frac{\partial T}{\partial y}.
\end{align}
\vspace{0.5cm}
Thus,
\begin{align}
S&=-\frac{C_1C_3+C_2C_4}{C_1^2+C_2^2},\label{60}\\[0.4em] N|\vec{B}|&=\frac{C_2C_3-C_1C_4}{C_1^2+C_2^2}.\label{61}
\end{align}

For the physical medium consisting of $u$ and $d$ quarks, the total currents are given as:
\begin{align}
J_x=&\sum_{a=u,d}\left[q_a(I_1)_aE_x+q_a(I_2)_aE_y+\beta (I_3)_a\frac{\partial T}{\partial x}+\beta(I_4)_a\frac{\partial T}{\partial y}\right]\\[0.4em]
J_y=&\sum_{a=u,d}\left[-q_a(I_2)_aE_x+q_a(I_1)_aE_y-\beta (I_4)_a\frac{\partial T}{\partial x}+\beta(I_3)_a\frac{\partial T}{\partial y}\right].
\end{align}
Setting the currents equal to 0 as earlier, we arrive at the Seebeck and Nernst coefficients of the composite medium:
\begin{align}
S&=-\frac{K_1K_3+K_2K_4}{K_1^2+K_2^2}\,,\label{64}\\[0.5em] N|\vec{B}|&=\frac{K_2K_3-K_1K_4}{K_1^2+K_2^2}.\label{65}
\end{align}
where, 
\begin{align}
K_1&=\sum_{a=u,d}q_a(I_1)_a\,,\qquad K_2=\sum_{a=u,d}q_a(I_2)_a\,,\nn\\[0.4em]
K_3&=\sum_{a=u,d}\beta(I_3)_a\,,\qquad \, K_4=\sum_{a=u,d}\beta(I_4)_a.\nn
\end{align}
\section{Quasiparticle description}
Quasiparticle description is a phenomenological description of quarks and gluons in a
thermal QCD medium, in which, thermal
masses of partons are generated, apart from their current masses in QCD Lagrangian.
These masses are generated due to the interaction of a
given parton with other partons in the medium, therefore, quasiparticle description describes the collective
properties of the medium. It can be applied to study several thermal properties of
QGP near the crossover temperature, $T_c$ , where perturbation theory cannot be used directly. Such a model was initially proposed by Goloviznin and Satz\cite{Goloviznin:ZPC57'1993}. Different versions of quasiparticle description exist in the literature based on different effective
theories, such as Nambu-Jona-Lasinio (NJL) model and its extension Polyakov-loop extended Nambu Jona Lasinio
model~\cite{Fukushima:PLB591'2004,Ghosh:PRD73'2006,Abuki:PLB676'2009}, 
Gribov-Zwanziger quantization \cite{Su:PRL114'2015,Florkowski:PRC94'2016}, thermodynamically consistent quasiparticle model \cite{Bannur:JHEP0709'2007}, 
etc. The results arrived at using these models suggest that it is possible to describe the high temperature QGP phase
by a thermodynamically consistent
quasiparticle model.
Our description relies on perturbative thermal
QCD, where the medium generated masses for quarks and
gluons are obtained from the poles of dressed propagators
calculated by the respective self-energies at finite temperature.

The relaxation time is infact an artifact of the quasiparticle description itself. The gluon exchange that takes place during parton scattering in the QGP medium is infested with infrared singularity owing to the zero rest mass of the gluon. The problem is however circumvented by the finite thermal mass acquired by the gluons (and also the quarks) in the quasiparticle description. This mass acts as an infrared cutoff in transverse gluon exchange processes that play the dominant role in bringing the system back to equilibrium post an infinitesimal disturbance. In deriving the relaxation-time, we take help of the Boltzmann transport equation via which we calculate initially the shear viscosity ($\eta$) of the medium.
\begin{equation}
\frac{1}{\eta}=\frac{\pi^7\alpha_s^2}{T^3\,480\,(\zeta(5))^2}\,
\mbox{ln}\left(\frac{T}{m_{\tiny gT}}\right).\label{Z9}
\end{equation}
The issue of the singularity in transverse gluon exchange processes can be clearly seen above, where, the inverse of shear viscosity would diverge in the absence of medium generated parton masses owing to the factor 
$\mbox{ln}\left(\frac{T}{m_{\tiny gT}}\right)$. After having calculated the shear viscosity, we associate with it, a viscous relaxation time $\tau$, defined by the relaxation time approximation of the Boltzmann transport equation. 
\begin{equation}
Df=-\frac{f(\vec{p})-f_0(\vec{p})}{\tau(\vec{p})},\label{z10}
\end{equation}
where, $D$ refers to the total time derivative. In conjunction with the definition of shear viscosity, this yields the relaxation time as follows:
\begin{align}
 \tau &\simeq \frac{\eta}{1.404T^4}\nonumber \\[0.5em] 
 &=\frac{N_f+0.6}{0.4(N_f+6)T\alpha_s^2}\,\frac{1}{\text{ln}(T/m_{gT})}. \label{tau}
\end{align}
In the quasiparticle description of quarks and gluons
in a thermal medium, all quark flavours
(with current/vacuum masses, $m_i << T$) 
acquire the same thermal mass~\cite{Braaten:PRD45'1992,Peshier:PRD66'2002}
\begin{equation}
m_T^2=\frac{g^2(T)T^2}{6},\label{Gluon mass}
\end{equation}
which is, however, modified in the presence of a finite chemical 
potential~\cite{Kakade:PRC92'2015}
\begin{equation}\label{Quark mass}
m_{T,\mu}^2=\frac{g^2(T)T^2}{6}\left(1+\frac{\mu^2}{\pi^2T^2}\right).
\end{equation}
We take the pure thermal ($B=0$) expressions of quasiparticle masses with magnetic field dependence coming in implicitly via the coupling constant. This is justified since we are working in a regime where $eB\ll T^2$. We use a one loop running coupling constant $\alpha_s (\Lambda^2,eB)$, 
which runs with both the magnetic field and temperature:\cite{Ayala:PRD98'2018}
\begin{equation}
\alpha_s(\Lambda^2,|eB|)=\frac{\alpha_s(\Lambda^2)}{1+b_1\alpha_s(\Lambda^2)\,\mbox{ln}\left(\frac{\Lambda^2}{\Lambda^2+|eB|}\right)},\label{coupling}
\end{equation}
\vspace{4mm}
where, $\alpha_s(\Lambda^2)$ is the one-loop running coupling in the absence of a magnetic field 
$$\alpha_s(\Lambda^2)=\frac{1}{b_1\,\mbox{ln}\left(\frac{\Lambda^2}{\Lambda_{QCD}^2}\right)},$$
with $b_1=(11N_c-2N_f)/12\pi$ and $\Lambda_{QCD}\sim 0.2$ GeV. The renormalisation scale is chosen 
to be $\Lambda=2\pi\sqrt{T^2+\frac{\mu^2}{\pi^2}}$. $\alpha_s(\Lambda^2,|eB|)$ determines $g(T,|eB|)$ via the relation
\begin{equation}
\alpha_s(\Lambda^2,|eB|)=\frac{g^2(T,|eB|)}{4\pi}.\label{cc}
\end{equation} 
Thus, the thermally generated mass takes on an implicit dependence on the magnetic field via the coupling constant.
We take the quasiparticle mass (squared) of $i$th 
flavor to be~\cite{Bannur:JHEP0709'2007,Bannur:PRC75'2007,Srivastava:PRD82'2010,Srivastava:PRD85'2012}:
\begin{equation}
m_{iT}^{\prime~2}=m_{i}^2+\sqrt{2}\,m_{i}\,m_T+m_T^2.\label{em}
\end{equation}

\section{Results}
The relaxation time used in this calculation is the one evaluated for a pure thermal medium ($B=0$) with a finite chemical potential with the justification being that temperature is the hardest scale in the problem. Also, the magnetic field dependence is implicit in the relaxation time via the coupling constant $\alpha_s$. The relaxation time for quarks, anti-quarks is given by \cite{Dey:PRD102'2020}:
\begin{equation}
\frac{1}{\tau (T,\mu)} = \frac{0.4.(N_f+6)}{(N_f+0.6)}~T\alpha_s^2\,
\mbox{ln}\left[\frac{1}{2\pi\alpha_s}\,\frac{1}{\left( (N_f+1)+
	\frac{3}{\pi^2} \sum_i\frac{\mu_i^2}{T^2}\right)}\right]\label{rt}.
\end{equation}
where, $N_f$ is the number of quark flavours in the medium and $\alpha_s$ is as given by Eq.\eqref{coupling}. 
\subsection{1-D Seebeck coefficient and its comparison with the strong magnetic field case}
\begin{figure}[H]
	\begin{subfigure}{0.48\textwidth}
		\includegraphics[width=0.95\textwidth]{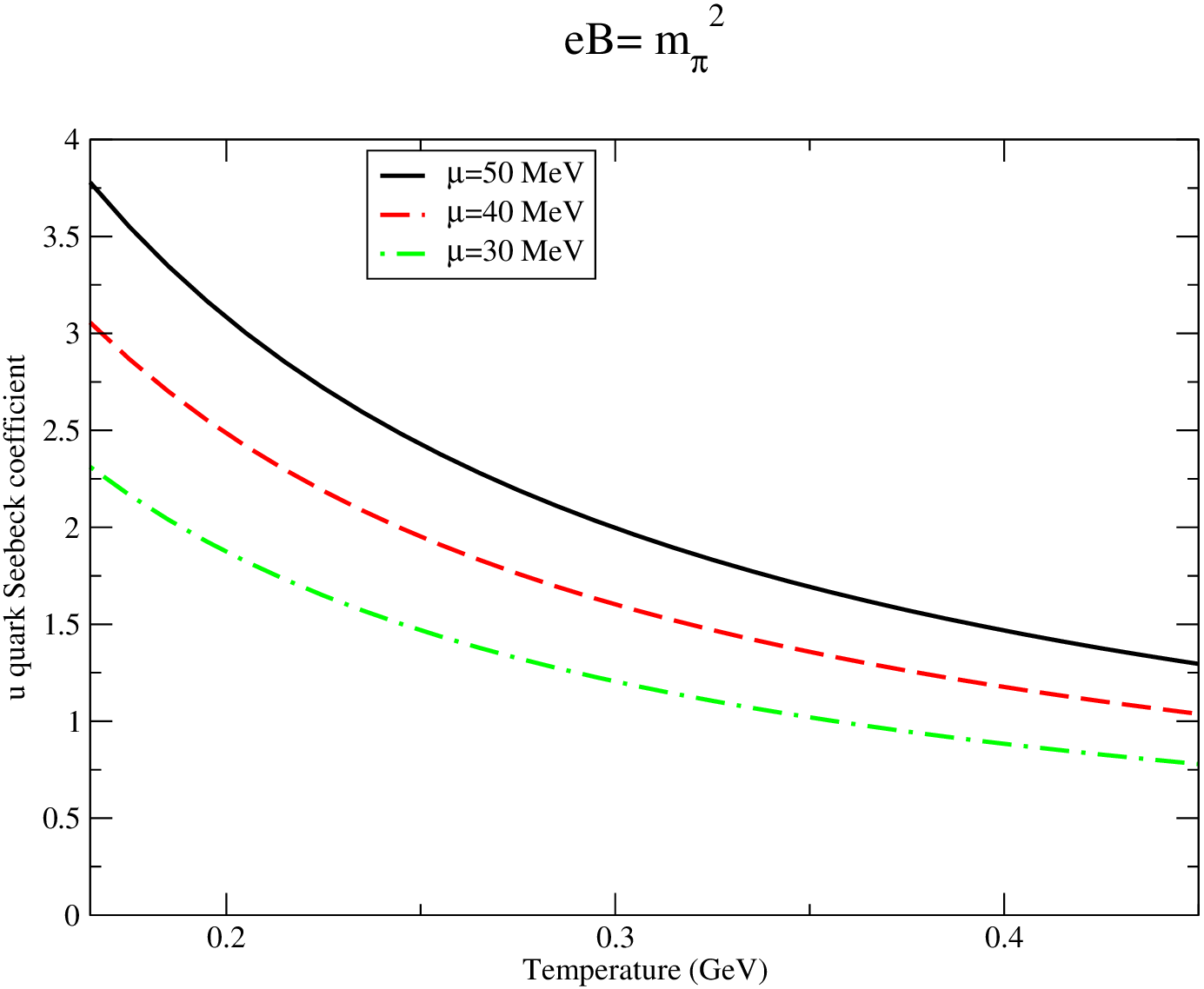}
		\caption{}\label{fig1a}
	\end{subfigure}
	\hspace*{\fill}
	\begin{subfigure}{0.48\textwidth}
		\includegraphics[width=0.95\textwidth]{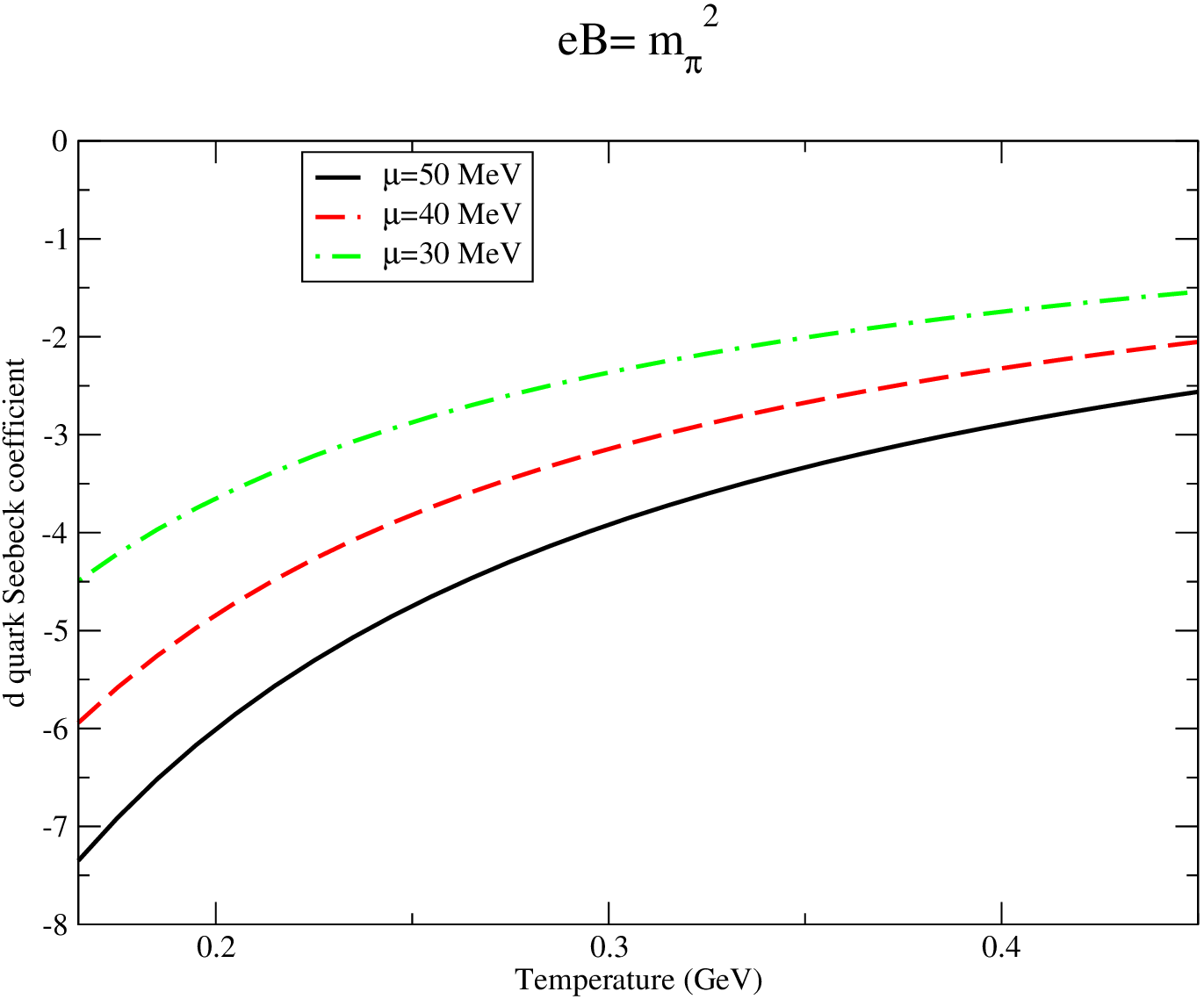}
		\caption{}\label{fig1b}
	\end{subfigure}
	\caption{Variation of Seebeck coefficient of 
		$u$ (a) and $d$ (b) quarks with 
		temperature for different fixed values of quark chemical potential.}\label{fig1}
\end{figure}
The variation of Seebeck coefficients of $u$ and $d$ quarks with temperature is shown in Fig.\eqref{fig1} for three different values of quark chemical potential in the presence of a weak ($B=0.06\,\mbox{GeV}^2$) magnetic field. We observe that the magnitude of Seebeck coefficient for both $u$ and $d$ quarks decreases with temperature. This is due to the fact that the net number density, ($n- 
\bar{n}$) (which is proportional to the net charge) decreases with the
temperature for a fixed $\mu$. The rate of increase (slope) is more pronounced at low temperatures as compared to higher temperatures. The coefficient increases with increasing chemical potential for a fixed value of temperature. For the $u$ quark, a finite chemical potential implies an abundance of positive charges (particles) over negative charges (anti-particles), leading to a greater thermoelectric current for higher chemical potential, and hence, a larger seebeck coefficient. Similarly, for the $d$ quark, a larger chemical potential means a larger abundance of negative charges over positive charges, leading to a more negative value of the Seebeck coefficient. The sign of the Seebeck coefficient is positive for the positively charged $u$ quark and negative for the negatively charged $d$ quark, in accordance with our expectation. 
\vspace{8mm}
\begin{figure}[H]
	\centering
	\includegraphics[width=9.2cm,height=8.8cm,keepaspectratio]{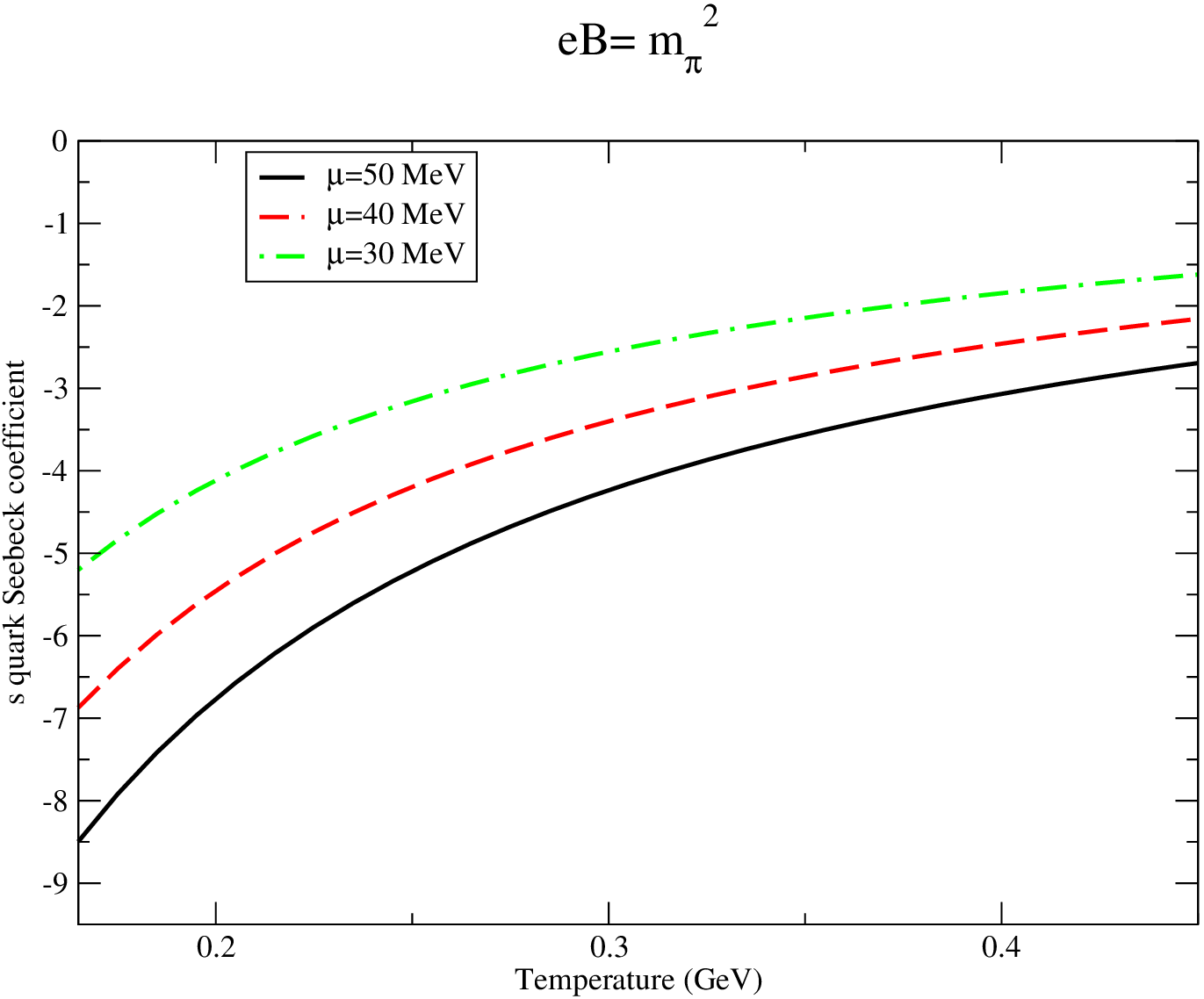}
	\caption{Variation of $s$ quark Seebeck 
		coefficient with 
		temperature for different fixed values of quark chemical potential.}
	\label{fig2}
\end{figure}
The Seebeck coefficient for $s$ quark exhibits the same trend as far as variation with temperature is concerned, as can be seen from Fig.(\ref{fig2}). For a given value of the magnetic field, the magnitude of the Seebeck coefficient decreases with increasing temperature. Also, the sign of the Seebeck coefficient is negative throughout the entire temperature range, as a consequence of the negative electric charge of the $s$ quark. Like in the case of $u$ and $d$ quarks, the magnitude of the $s$ quark Seebeck coefficient increases with increasing chemical potential. The $d$ and $s$ quarks carry the same electric charge. As such, the major differentiators of the two species are the strangeness quantum number and the current mass, with only the latter being relevant in this discussion. Comparing the Seebeck coefficients of the $d$ and $s$ quarks therefore gives us an idea as to how the mass of the particle affects the individual Seebeck coefficient. It can be seen from the comparison that a larger current quark mass has an amplifying effect on the seebeck coefficient in the case of a weak magnetic field. To clearly visualize this, we have plotted the individual Seebeck coefficient as a function of the current quark mass at different fixed values of temperature, taking the charge to be $-\frac{2}{3}e$, in the presence of a weak magnetic field in Fig.\eqref{fig6a}. As can be seen, the coefficient magnitude increases with the current mass, with the increase being more pronounced at lower temperatures.

To evaluate the total Seebeck coefficient of the medium comprising of multiple quark flavours ($u$, $d$, $s$ in our case), we make use of the individual Seebeck coefficients already obtained, and substitute in Eq.\eqref{47}.
\vspace{4mm}
\begin{figure}[H]
	\centering
	\includegraphics[width=9.2cm,height=8.6cm,keepaspectratio]{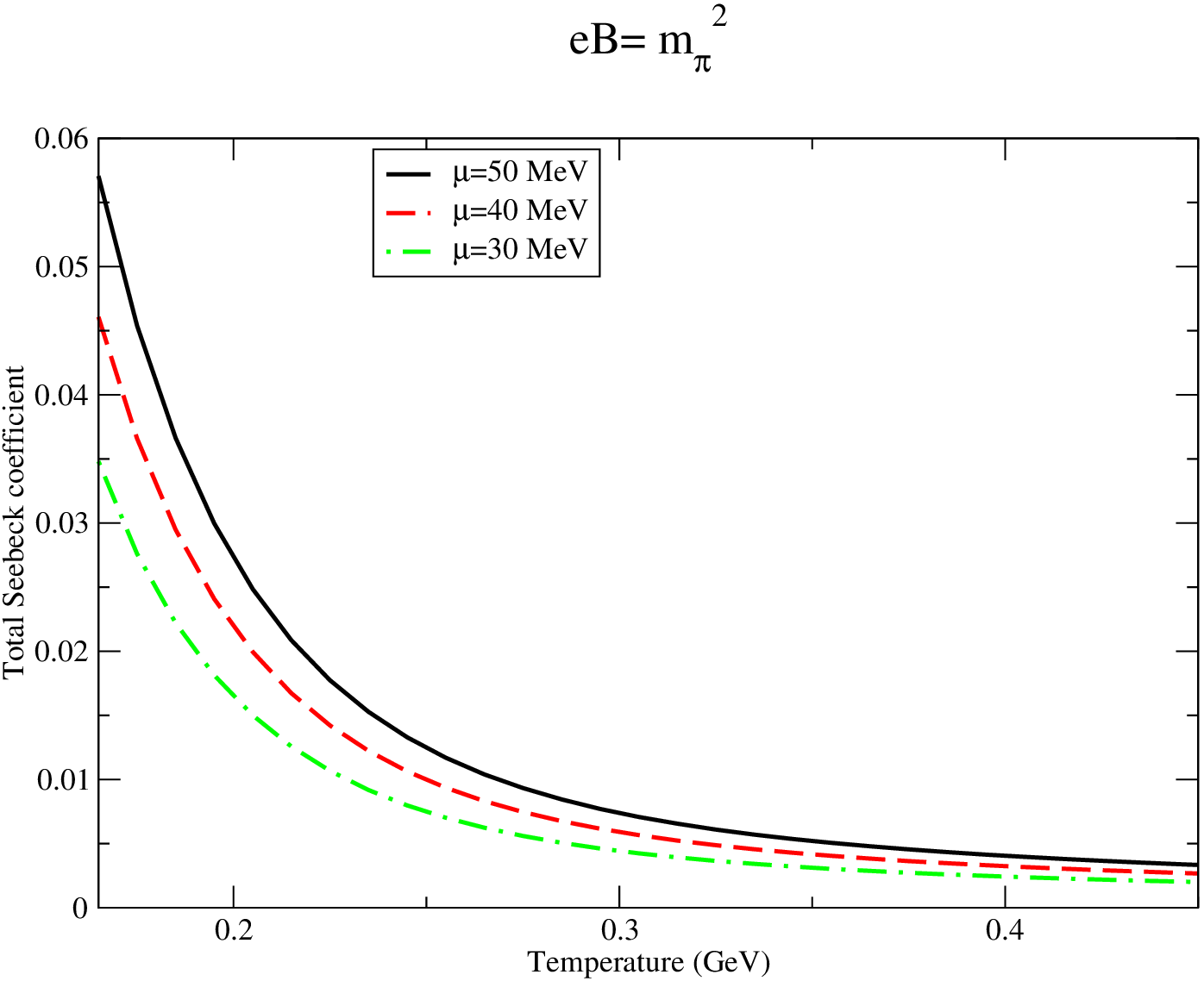}
	\caption{Variation of total Seebeck 
		coefficient with
		temperature for different fixed values of quark chemical potential.}
	\label{fig3}
\end{figure}
As can be seen from the Fig.\eqref{fig3}, the total Seebeck coefficient of the medium decreases with increasing temperature. Similar to the case of single species, the magnitude of the coefficient increases with increasing chemical potential. As mentioned earlier, the total Seebeck coefficient is a weighted average of the individual Seebeck coefficients. Although the individual coefficients for $d$ and $s$ quarks are negative, the weighted average renders the total Seebeck coefficient positive for the entire temperature range. \\\\
 We now discuss briefly the procedure for calculating the Seebeck coefficient in the presence of a strong magnetic field\cite{Dey:PRD102'2020}.
 Charged particles in the presence of a magnetic field occupy discrete energy levels ($n=0,1,2,\cdots$). This is referred to as Landau quantization and is thus applicable for quarks in a magnetised QGP\cite{Bruckmann:PRD96'2017}. The thermal occupation probability of higher landau levels by the quarks is found to be exponentially suppressed by $e^{-\frac{\sqrt{eB}}{t}}$. Thus for strong magnetic fields ($\sqrt{eB}\gg T$), the dominant contribution comes from the lowest Landau level ($n=0$) as the higher ones do not contribute to transport phenomena in leading order and can be neglected in calculations. This is the Lowest Landau Level (LLL) approximation. Consequently, the quark dispersion relation 
\begin{equation}
\omega_{(n)}(p_L)=\sqrt{p_{L}^2+m^2+2n|qB|},
\end{equation}
gets reduced to
\begin{equation}
\omega(p_L)=\sqrt{p_{L}^2+m^2},\label{dr}
\end{equation}
where, $m$, $q$ are respectively the mass and charge of the quark and $p_L$ is the component of quark momentum along the direction of $\vec{B}$. The transverse momentum (squared) $2n|qB|$ vanishes in the LLL approximation. Thus we have a dimensional reduction $D\longrightarrow D-2$ in fermion dynamics in a strong magnetic field\cite{Gusynin:NuclPhysB462'1996}. Following this approach, the effect of strong magnetic field is incorporated in the evaluation of the Seebeck coefficient. Further, instead of Eq.\eqref{Quark mass}, the thermomagnetic quark mass- which is nothing but the pole of the full quark propagator in a strong magnetic field, is used\cite{Rath:PRD100'2019}
\begin{equation}
m_{T,B}^2 =\frac{g^2|qB|}{3\pi^2}\left[\frac{\pi T}{2m}-\ln(2)\right],
\end{equation}
where, $m$ and $q$ are the current mass and electric charge of the quark in question with $g$ obtained from Eq.\eqref{cc}. As earlier, the effective quasiparticle mass (squared) of $i$th flavor is taken to be
\begin{equation}
m_{i(T,B)}^{\prime~2}=m_{i}^2+\sqrt{2}\,m_{i}\,m_{i(T,B)}+m_{i(T,B)}^2,\label{emb}
\end{equation}
so that Eq.\eqref{dr} for $i$-th quark becomes
 \begin{equation}
 \omega_i(p_L)=\sqrt{p_{L}^2+m_{i(T,B)}^{\prime~2}}.\label{59}
 \end{equation}
 Starting from the relativistic Boltzmann transport equation [Eq.\eqref{2}], the final expression for the individual Seebeck coefficient in the presence of a strong magnetic field comes out to be 
 \begin{equation}
 S=\frac{1}{2Tq}\frac{H_{1}}{H_{2}},
 \end{equation}
 where,
 \begin{eqnarray}
 H_1&=&\int \frac{dp_z} {w^2}\,\tau_{B}\,p_z^2\left\{-\bar{f}
 (1-\bar{f})(\omega+\mu)+f(1-f)(\omega-\mu)
 \right\},\\
 H_2&=&\int \frac{dp_z}{w^2}\,\tau_{B}\,p_z^2\left\{\bar{f}(1-\bar{f})+f(1-f)
 \right\}.
 \end{eqnarray}
 Here, $f$ and $\bar{f}$ denote the equilibrium distribution functions for the concerned quark and its antiquark respectively, $p_z\equiv p_L$ with $\omega$ being given by Eq.\eqref{59}. $\tau_B$ denotes the relaxation-time 
 for quarks in the presence of strong magnetic field, which, in 
 the Lowest Landau Level (LLL) approximation 
 is given by\cite{Hattori:PRD95'2017}:
 \begin{equation}
 \tau_B(T,B)=\frac{w\left(e^{\beta \omega}
 	-1\right)}{\alpha_s \left(\Lambda^2, eB\right) C_2\,m^2\left(e^{\beta 
 		\omega}+1\right)}\left[\frac{1}{\int dp'^3 
 	\frac{1}{w'\left(e^{\beta \omega'}+1
 		\right)}}\right],\label{rtb}
 \end{equation}
 where, $C_2=4/3$ is the Casimir factor, $m$ is the mass of the concerned quark species and $\alpha_s (\Lambda^2,eB)$ is given by Eq.\eqref{cc}. Proceeding in a similar fashion, the total Seebeck coefficient of the composite medium is given by
 \begin{equation}
 S=\frac{1}{2T}\frac{\sum_i q_i|q_iB|(H_1)_i}
 {\sum_i q_i^2|q_iB|(H_2)_i},\label{tc}
 \end{equation} 
 which could be further expressed in terms of the weighted average of 
 individual Seebeck coefficients:
 \begin{equation}
 S_{\text{total}}=\frac{\sum_iS_i|q_i|^3(H_2)_i}{\sum_i|q_i|
 	^3(H_2)_i},\label{tc1}
 \end{equation}
 where, the summation is over quark flavours present in the medium and $S_i$ denotes the individual Seebeck coefficient of a medium containing $i$-th quark flavour alone.
\vspace{6mm}
\begin{figure}[H]
	\begin{subfigure}{0.48\textwidth}
		\includegraphics[width=0.95\textwidth]{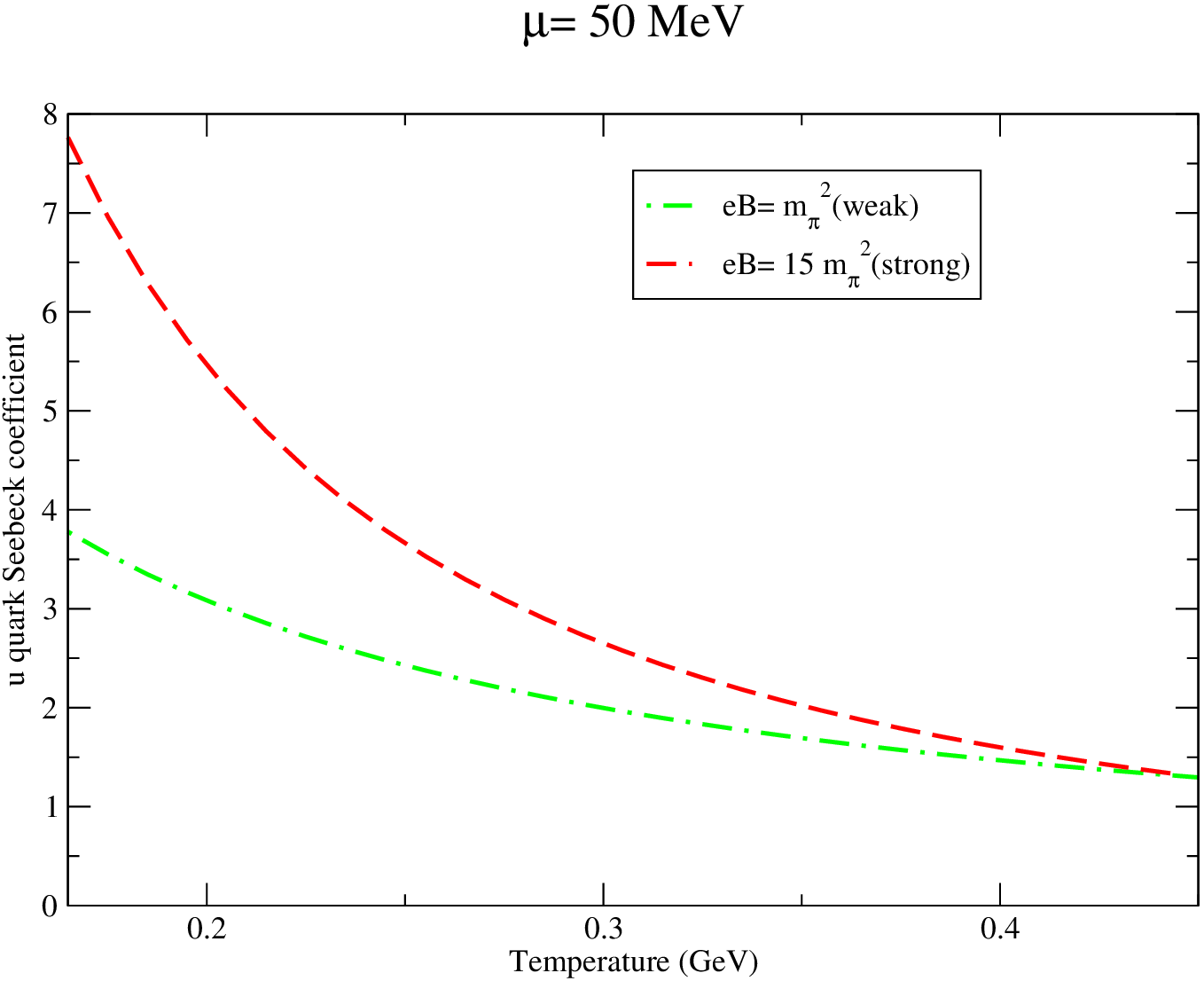}
		\caption{}\label{fig4a}
	\end{subfigure}
	\hspace*{\fill}
	\begin{subfigure}{0.48\textwidth}
		\includegraphics[width=0.95\textwidth]{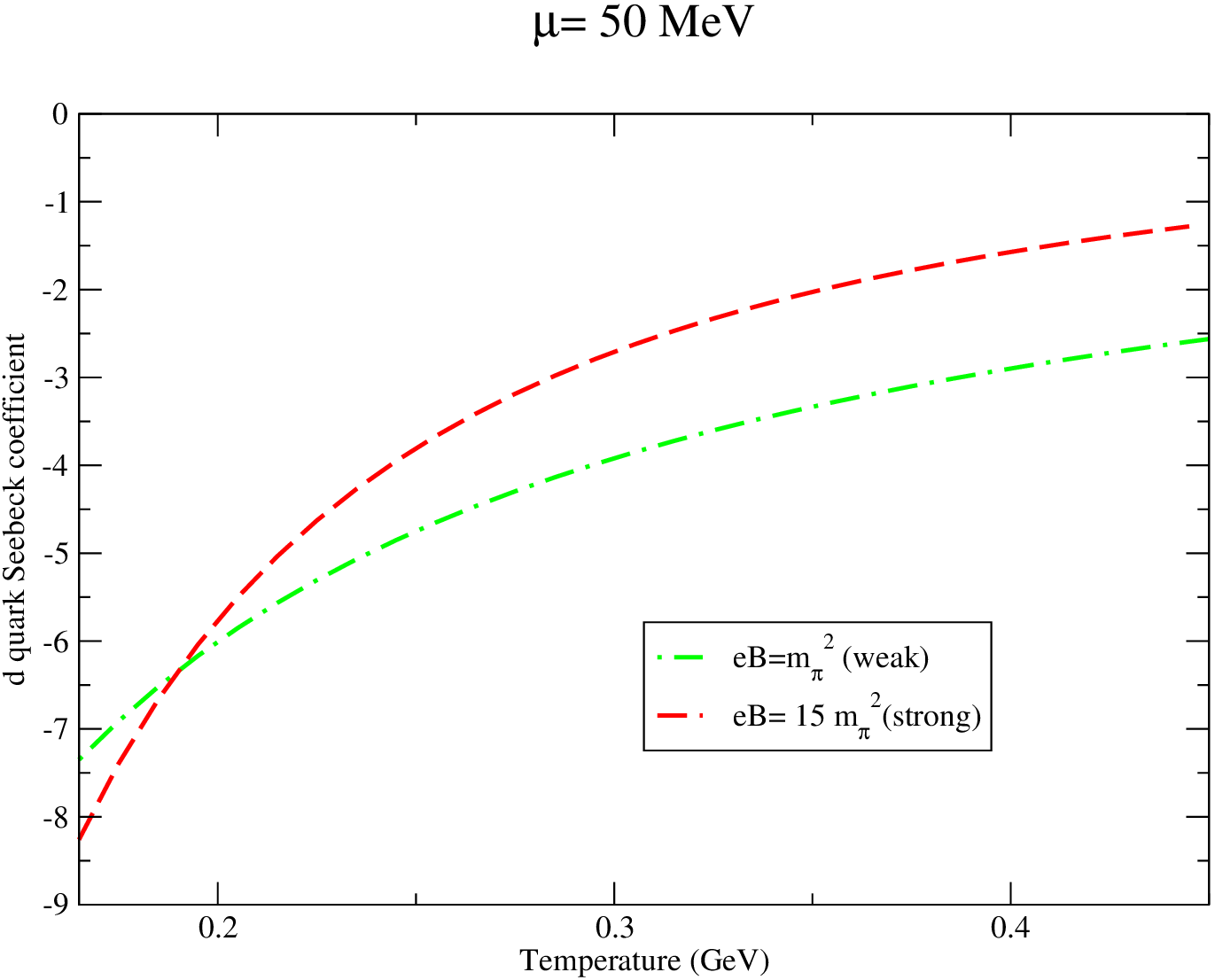}
		\caption{}\label{fig4b}
	\end{subfigure}
	\caption{Variation of Seebeck coefficient of 
		$u$ (a) and $d$ (b) quarks with 
		temperature for different fixed values of magnetic field.}\label{fig4}
	\end{figure}
\vspace{-2mm}
In Fig.\eqref{fig4a} and \eqref{fig4b}, we have drawn a comparison among the Seebeck coefficients of the $u$ and $d$ quarks, respectively, in different domains of magnetic field strengths. The behaviour is similar in some aspects for both the light quarks in that the magnitudes of the respective Seebeck coefficients decrease with temperature for both strong and weak magnetic field. For the $u$ quark, the Seebeck coefficient is significantly larger near $T_c$ in the strong magnetic field case, compared to the weak $B$ case. However, at higher temperatures ($\sim 3T_c$), the Seebeck coefficients for both the strong and weak $B$ cases converge.
For the $d$ quark, the magnitude of the Seebeck coefficient is again maximum in the presence of strong magnetic field near $T_c$. However, at higher temperatures, the trend is reversed and the Seebeck coefficient magnitude is greater for the weak $B$ case. So, the variation of individual Seebeck coefficient with temperature or the temperature sensitivity is maximum in the presence of a strong magnetic field which holds true for both the light quarks. 

\vspace{8mm}
\begin{figure}[H]
	\centering
	\includegraphics[width=9.2cm,height=8.8cm,keepaspectratio]{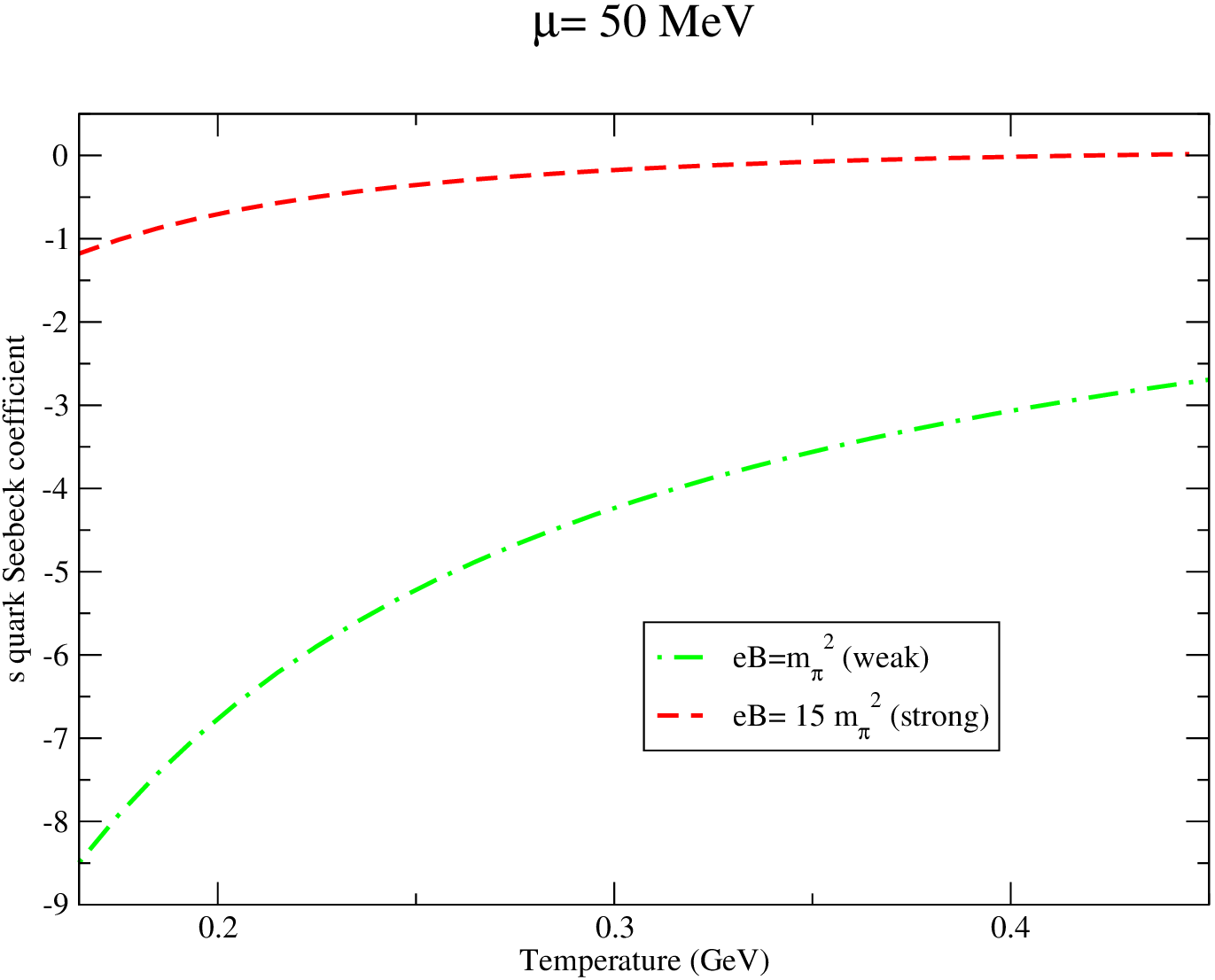}
	\caption{Variation of $s$ quark Seebeck 
		coefficient with 
		temperature for different fixed values of magnetic field.}
	\label{fig5}
\end{figure}
Fig.\eqref{fig5} compares the variation of $s$ quark Seebeck coefficient with temperature at different magnetic field strengths. The usual trends of decrease of the Seebeck coefficient magnitude with temperature and increase with chemical potential, hold. As earlier, the rate of increase is more pronounced at lower temperatures. However, the hierarchy of Seebeck coefficient magnitudes is reversed, with the magnitude being greater in the presence of weak magnetic field compared to that in the presence of strong magnetic field in the entire temperature range. We can analyse the effect of current quark mass on the magnitude of individual coefficients at different field strengths by comparing Fig.\eqref{fig4b} and Fig.\eqref{fig5}. It is clear that in the presence of a strong magnetic field, a greater current mass of the particle (quark) suppresses the ability of the medium to convert a temperature gradient into electric current whereas if the magnetic field is weak, the coefficient gets enhanced slightly for the quark with the larger current mass. The same comparison can also shed light on temperature sensitivity of individual coefficients as a function of particle current mass for different field strengths. In the case of light quarks [Fig.\eqref{fig4a}, Fig.\eqref{fig4b}], it was observed that the variation of Seebeck coefficient with temperature, and thereby the range of the Seebeck coefficient in the given temperature range, was most pronounced in the presence of strong magnetic field. For the $s$ quark, however, the variation is far less pronounced in the presence of strong magnetic field compared to when the magnetic field is weak. This can be seen clearly in Fig.\eqref{fig6b}, where we have plotted the range of individual Seebeck coefficient $\left(|S_{T_{\text{max}}}-S_{T_{\text{min}}}|\right)$ as a function of the current quark mass for both the weak $B$ and strong $B$ cases. The opposite trends are clearly visible, based on which, it can be argued that a larger current mass of the particle decreases the temperature sensitivity of the individual Seebeck coefficient when the background magnetic field is strong, whereas it causes a slight increase in the temperature sensitivity when the magnetic field is weak.  In Fig.\eqref{fig6a}, we have shown the dependence of individual Seebeck coefficient with $q=-2e/3$ and $\mu=50$ MeV on the current quark mass.

\vspace{6mm}
\begin{figure}[H]
	\begin{subfigure}{0.48\textwidth}
		\includegraphics[width=0.95\textwidth]{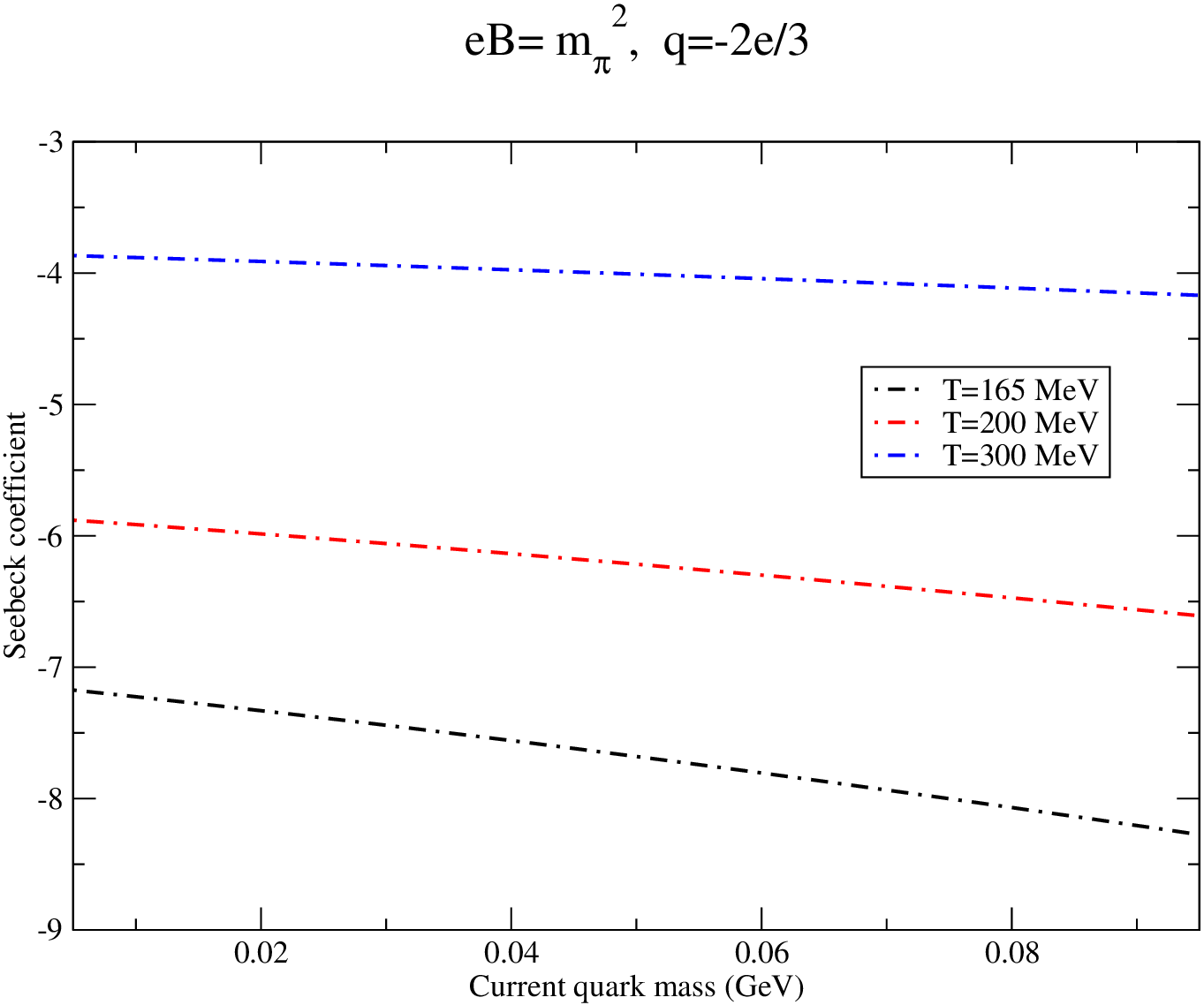}
		\caption{}\label{fig6a}
	\end{subfigure}
	\hspace*{\fill}
	\begin{subfigure}{0.48\textwidth}
		\includegraphics[width=0.95\textwidth]{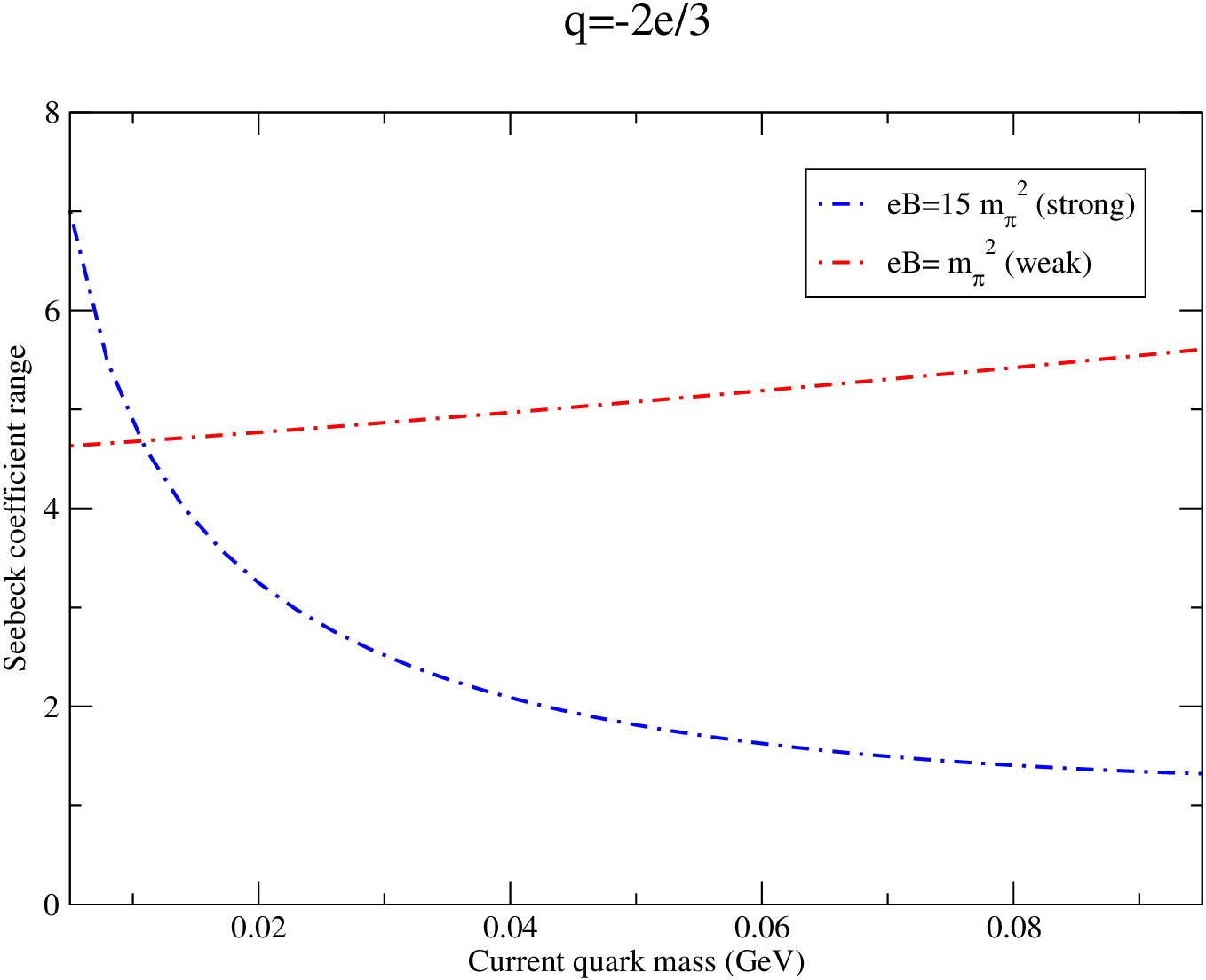}
		\caption{}\label{fig6b}
	\end{subfigure}
	\caption{\textbf{Left panel:} Variation of individual Seebeck coefficient as a function of current quark mass for a weak background magnetic field at different fixed values of temperature with $q=-\frac{2}{3}e$ and $\mu=50$ MeV. \textbf{Right panel:} Variation of range of individual Seebeck coefficient (Temperature sensitivity) as a function of current quark mass for different strengths of the background magnetic with $q=-\frac{2}{3}e$ and $\mu=50$ MeV.} \label{fig6}
\end{figure}

\begin{figure}[H]
	\centering
	\includegraphics[width=9.2cm,height=6.4cm,keepaspectratio]{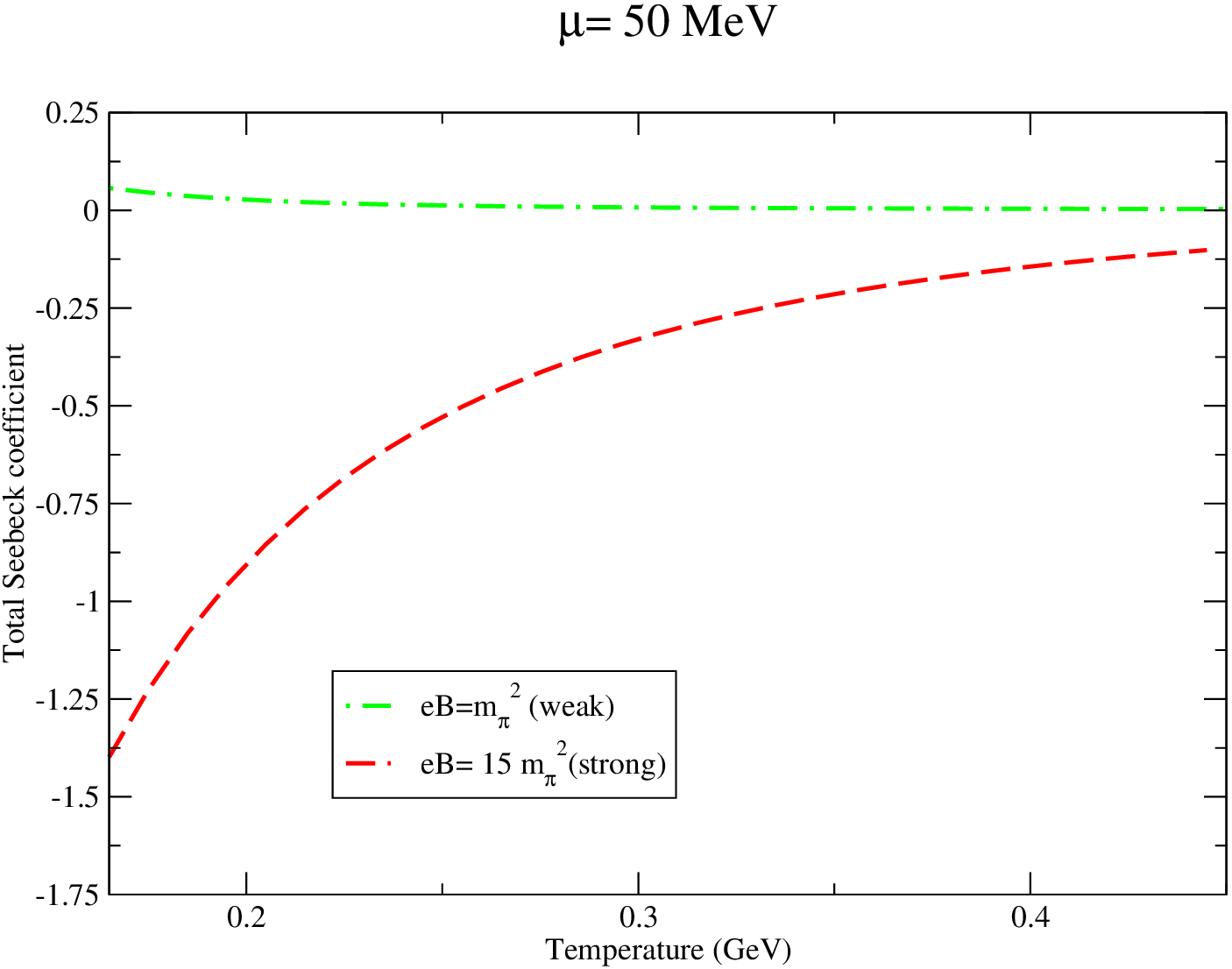}
	\caption{Variation of total Seebeck 
		coefficient of the composite medium with
		temperature for different fixed values of magnetic field.}
	\label{fig7}
\end{figure}
Fig.\eqref{fig7} shows the variation of the total Seebeck coefficient of the composite 
medium with temperature for different magnetic field strengths. As can be seen, 
the total Seebeck coefficient has a very small positive value for the weak magnetic field case for which the variation with temperature is 
very feeble in the entire temperature range. It should be remembered that the thermoelectric responses
of positive and negative charges ($u$ and $d$ quarks in our work) are competitive and since the total 
Seebeck coefficient of the medium in the framework adopted here 
is a weighted average of the individual coefficients,
the former could be quite different from the latter values. This is reflected  
in the strong $B$ case, where, the total Seebeck coefficient is negative in the entire temperature range, although the individual coefficients are still positive for positive charges and negative for negative ones. Physically, this implies that the induced electric field is generated 
opposite to the temperature gradient and the magnitude decreases with increasing temperature.
From Fig.\eqref{fig7} 
it is clear that 
in the composite medium, the thermoelectric response is the most 
effective in the presence of a strong magnetic field where the direction of the induced electric 
field is opposite to the direction of the temperature gradient. The response in the case of weak $B$ is comparatively feeble and monotonous in the entire temperature range.
\subsection{Seebeck and Nernst coefficients from 2-D formulation} 
\begin{figure}[H]
	\begin{subfigure}{0.48\textwidth}
		\includegraphics[width=0.95\textwidth]{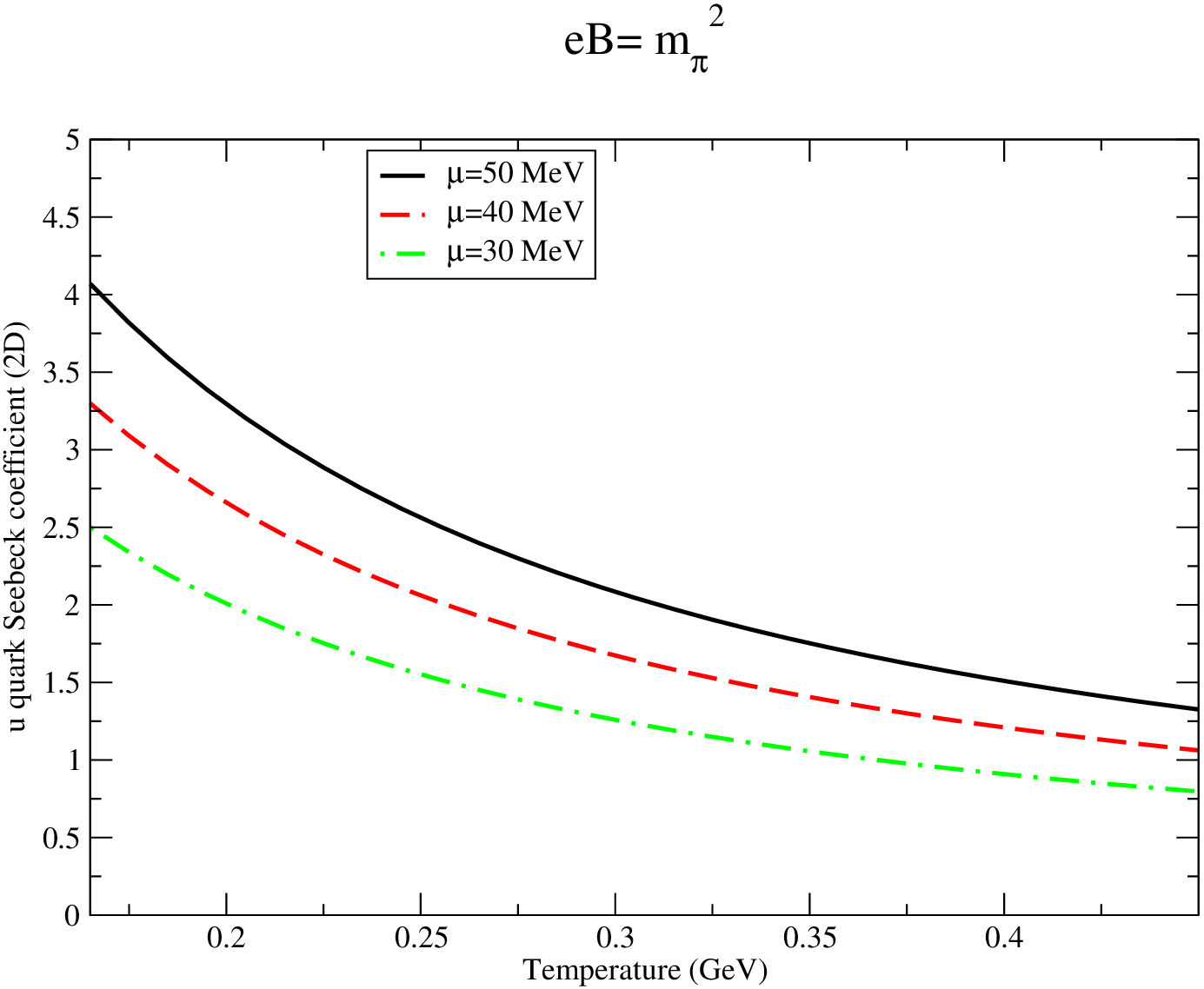}
		\caption{}\label{fig8a}
	\end{subfigure}
	\hspace*{\fill}
	\begin{subfigure}{0.48\textwidth}
		\includegraphics[width=0.95\textwidth]{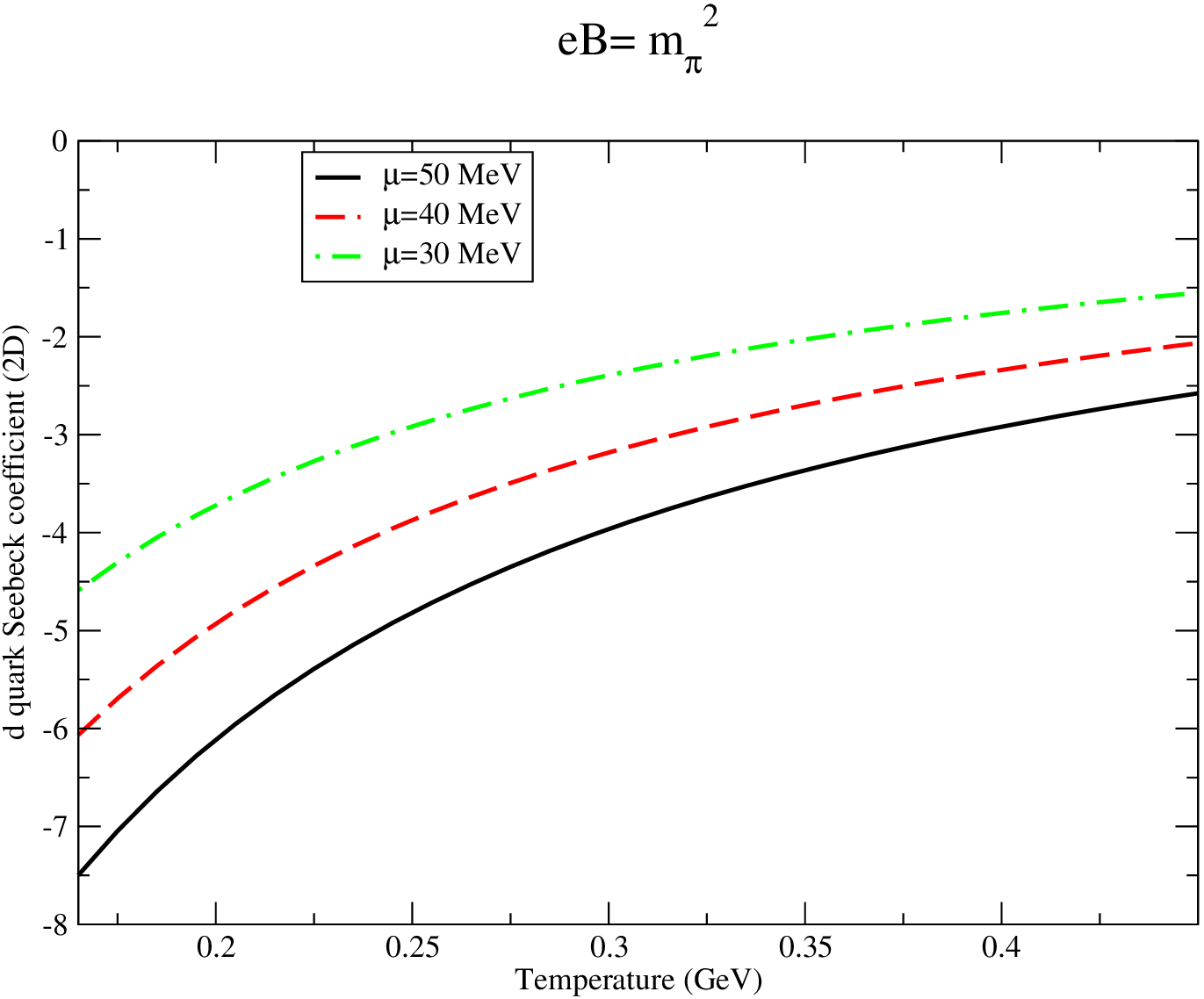}
		\caption{}\label{fig8b}
	\end{subfigure}
	\caption{Variation of Seebeck coefficient of 
		$u$ (a) and $d$ (b) quarks with 
		temperature for different fixed values of quark chemical potential.}\label{fig8}
\end{figure}
As can be seen in Fig.\eqref{fig8a} and \eqref{fig8b}, the magnitude of the individual coefficients decreases with temperature and increases with chemical potential, similar to the observations in the 1-D result. The individual Seebeck coefficients are ratios of two integrals [Eq.\eqref{60}]. The numerator and denominator of Eq.\eqref{60} for both $u$ and $d$ quarks  are all increasing functions of temperature as far as the absolute values are concerned. However, the ratios are monotonically decreasing functions of temperature for the temperature range considered here. The magnitudes and range of the coefficients also bear a close resemblance to the 1-D results. As expected, the coefficient is positive for the positively charged $u$ quark and negative for the negatively charged $d$ quark. The magnitudes of numerators and denominators in Eq.\eqref{60} for both $u$ and $d$ quarks are increasing functions of chemical potential as well. However, the rate of increase is more pronounced for the numerator than the denominator. This explains the overall increase of the individual coefficients with chemical potential.  
\begin{figure}[H]
	\begin{subfigure}{0.48\textwidth}
		\includegraphics[width=0.95\textwidth]{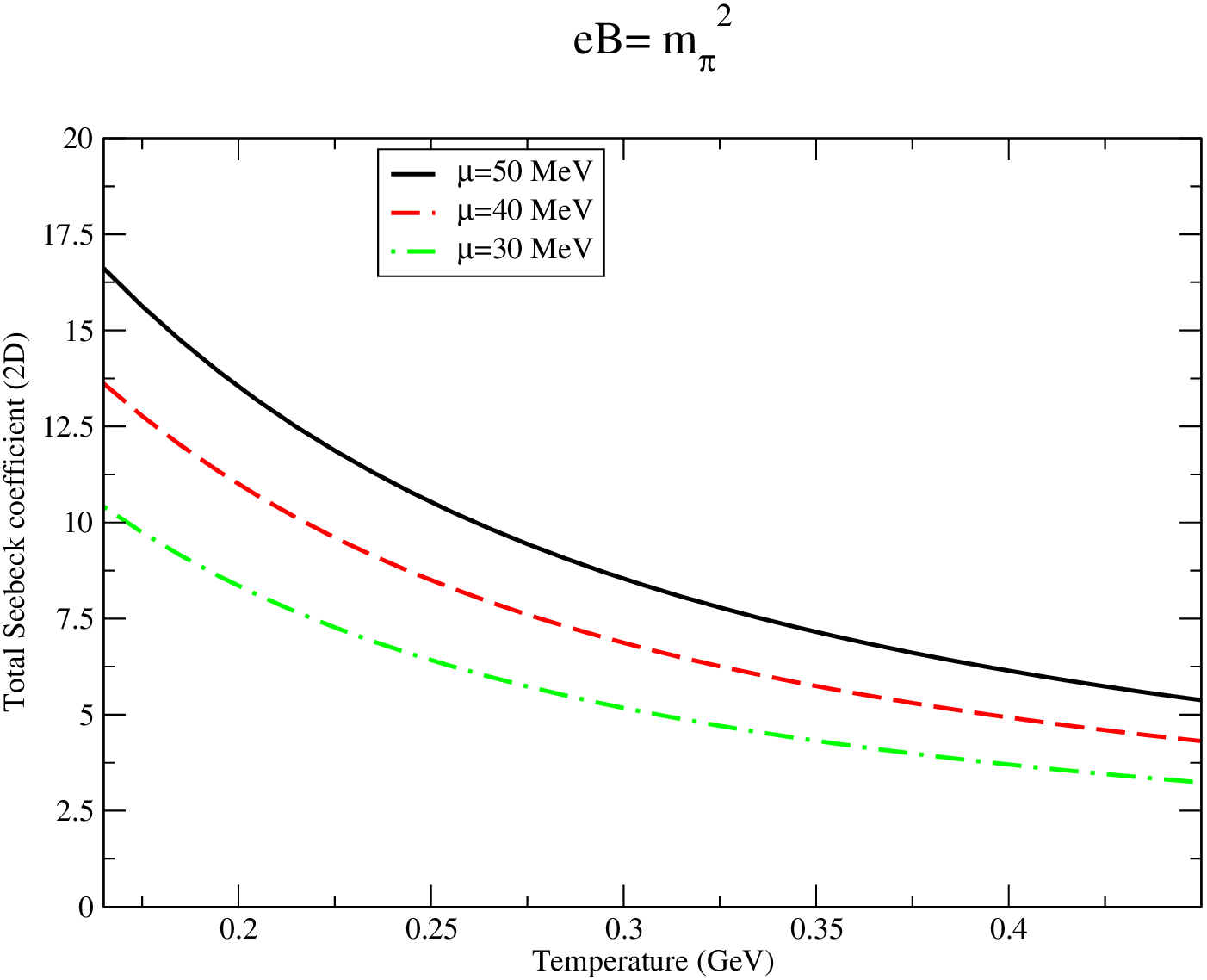}
		\caption{}\label{fig9a}
	\end{subfigure}
	\hspace*{\fill}
	\begin{subfigure}{0.48\textwidth}
		\includegraphics[width=0.95\textwidth]{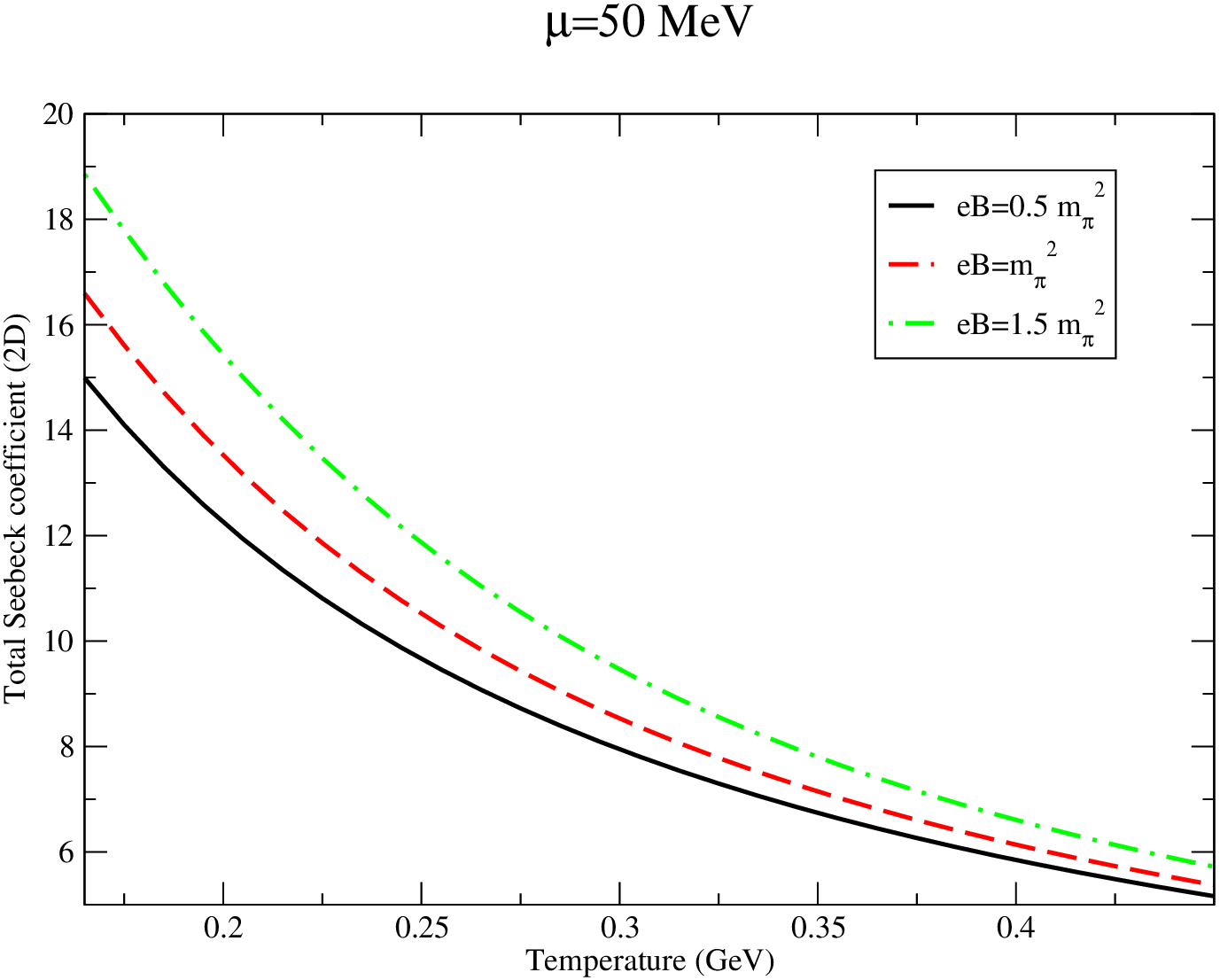}
		\caption{}\label{fig9b}
	\end{subfigure}
	\caption{\textbf{Left panel:} Variation of total Seebeck 
		coefficient of the composite medium with
		temperature at $eB=m_{\pi}^2$ for different fixed values of chemical potential. \textbf{Right panel:} Variation of total Seebeck coefficient with temperature at $\mu=50$ MeV for different fixed values of magnetic field.}\label{fig9}
\end{figure}

\begin{figure}[H]
	\begin{subfigure}{0.48\textwidth}
		\includegraphics[width=0.95\textwidth]{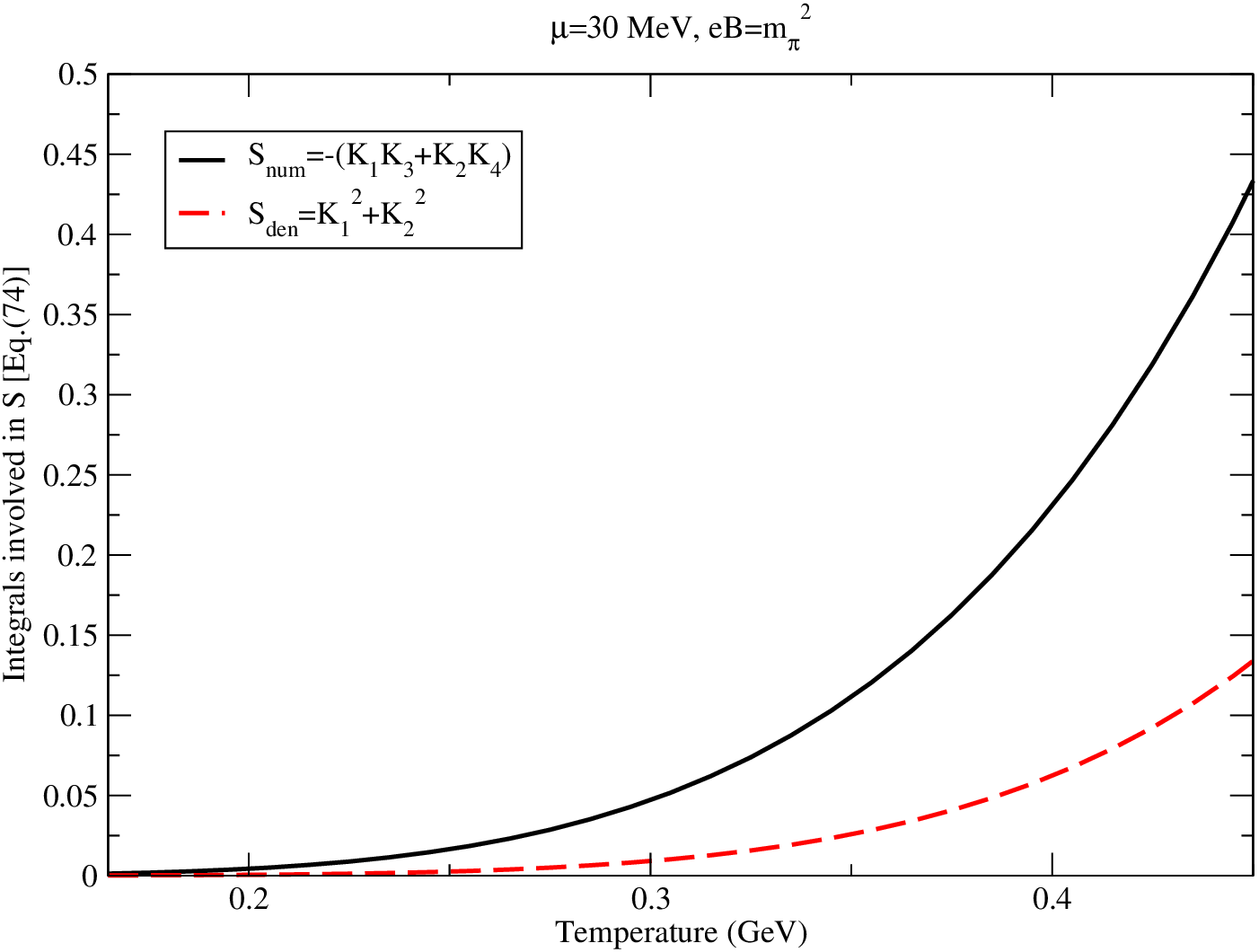}
		\caption{}\label{fig10a}
	\end{subfigure}
	\hspace*{\fill}
	\begin{subfigure}{0.48\textwidth}
		\includegraphics[width=0.95\textwidth]{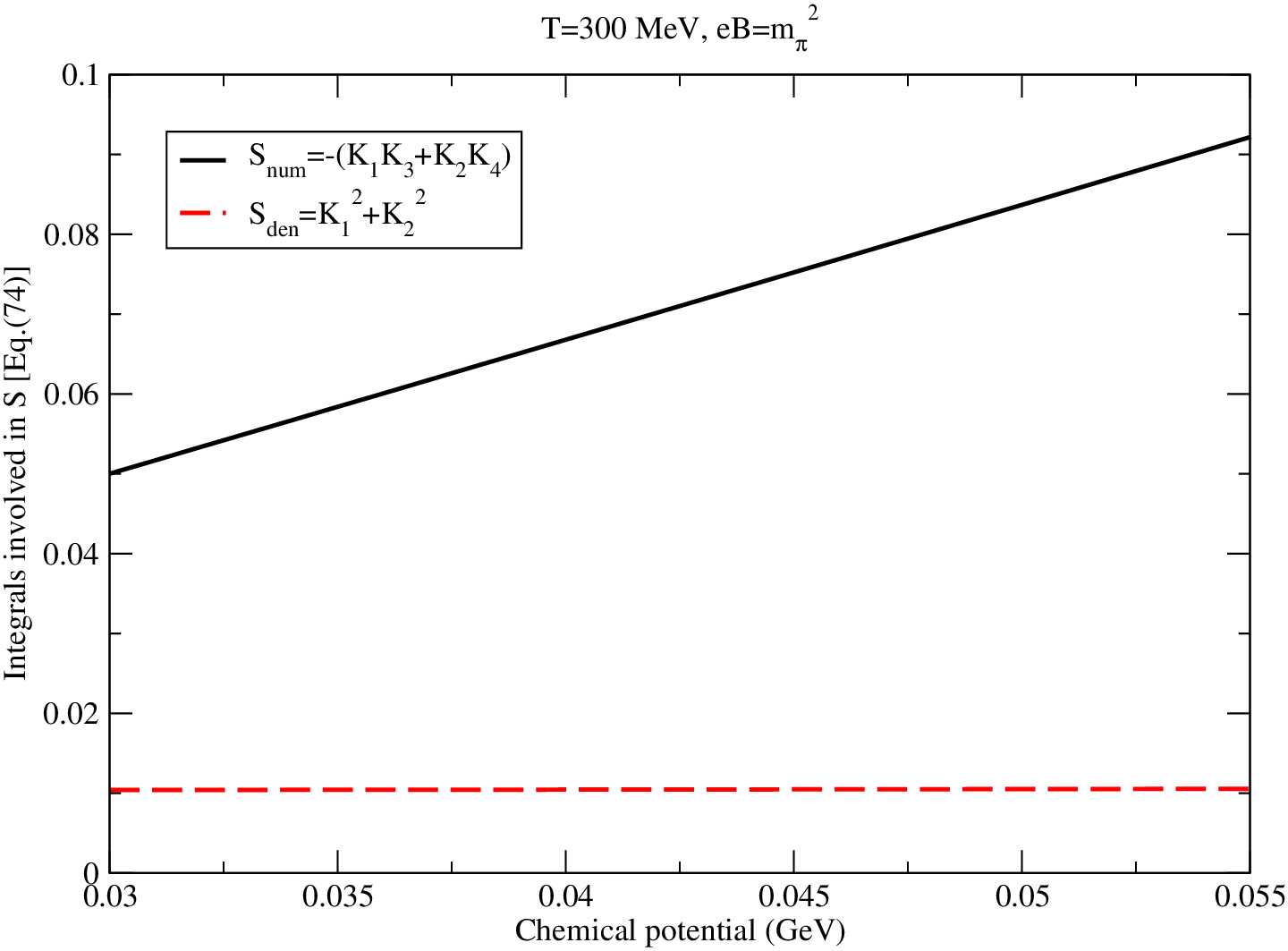}
		\caption{}\label{fig10b}
	\end{subfigure}
	\caption{\textbf{Left panel:} Variation of numerator and denominator of total Seebeck coefficient [Eq.\eqref{64}] with temperature for $eB=m_{\pi}^2$, $\mu=30$ MeV. \textbf{Right panel:} Variation of numerator and denominator of total Seebeck coefficient [Eq.\eqref{64}] with chemical potential for $eB=m_{\pi}^2$, $T=300$ MeV.}\label{fig10}
\end{figure}
Fig.\eqref{fig9a} shows the variation of Seebeck coefficient of the composite medium composed of $u$ and $d$ quarks with temperature. We see the earlier trend of decrease of coefficient magnitude with temperature. Also, the coefficient is positive and increases with increasing chemical potential. These trends can be understood from analysing the integrals in the numerator and denominator of Eq.\eqref{64}. Fig.\eqref{fig10a} shows the variation of the numerator and denominator integrals in Eq.\eqref{64} with temperature. It can be seen that both the numerator and the denominator are increasing functions of temperature. The comparative increase is such that the ratio is rendered a decreasing function of temperature. The variation with chemical potential [Fig.\eqref{fig10b}] clearly shows that the increase in the value of the numerator integral with temperature is significantly more pronounced than that of the denominator integral. This explains the rise in coefficient magnitude with increasing chemical potential. 
It should be noted that the values of the coefficient in the entire temperature range is about 2 orders of magnitude greater than the 1-D result. In this regard, it is important to note that unlike Eq.\eqref{tc1}, there exists no simple way to write the total Seebeck coefficient in the 2-D formulation as a weighted average of individual Seebeck coefficients 
 This suggests that the evaluation of Seebeck coefficient depends on the methodology adopted to evaluate it. Fig.\eqref{fig9b} shows the effect of magnetic field on the temperature dependence of the total Seebeck coefficient. As can be seen, the total Seebeck coefficient increases with increasing background magnetic field and records the same decreasing trend with temperature for each value of the magnetic field.  
\begin{figure}[H]
	\begin{subfigure}{0.48\textwidth}
		\includegraphics[width=0.95\textwidth]{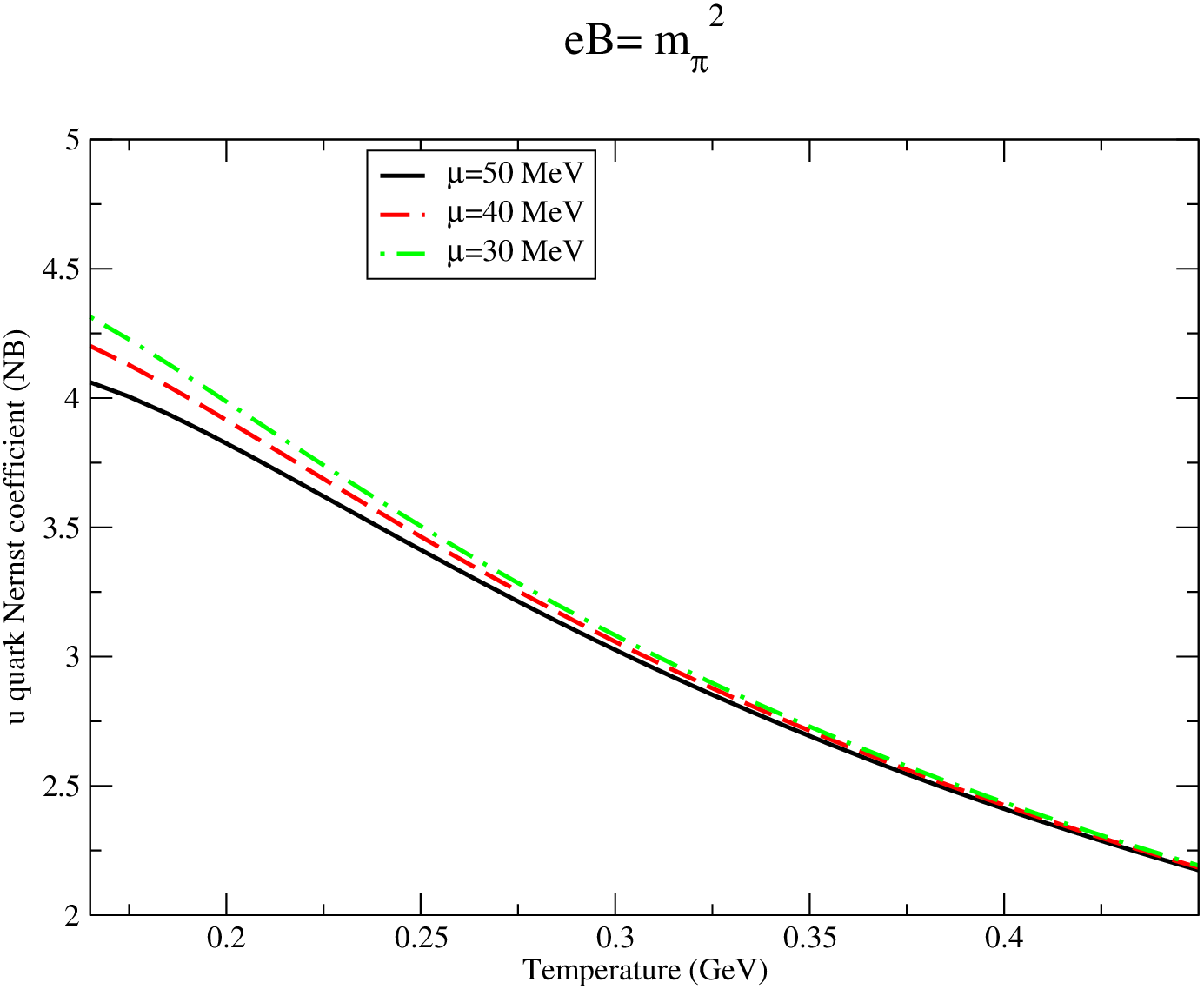}
		\caption{}\label{fig11a}
	\end{subfigure}
	\hspace*{\fill}
	\begin{subfigure}{0.48\textwidth}
		\includegraphics[width=0.95\textwidth]{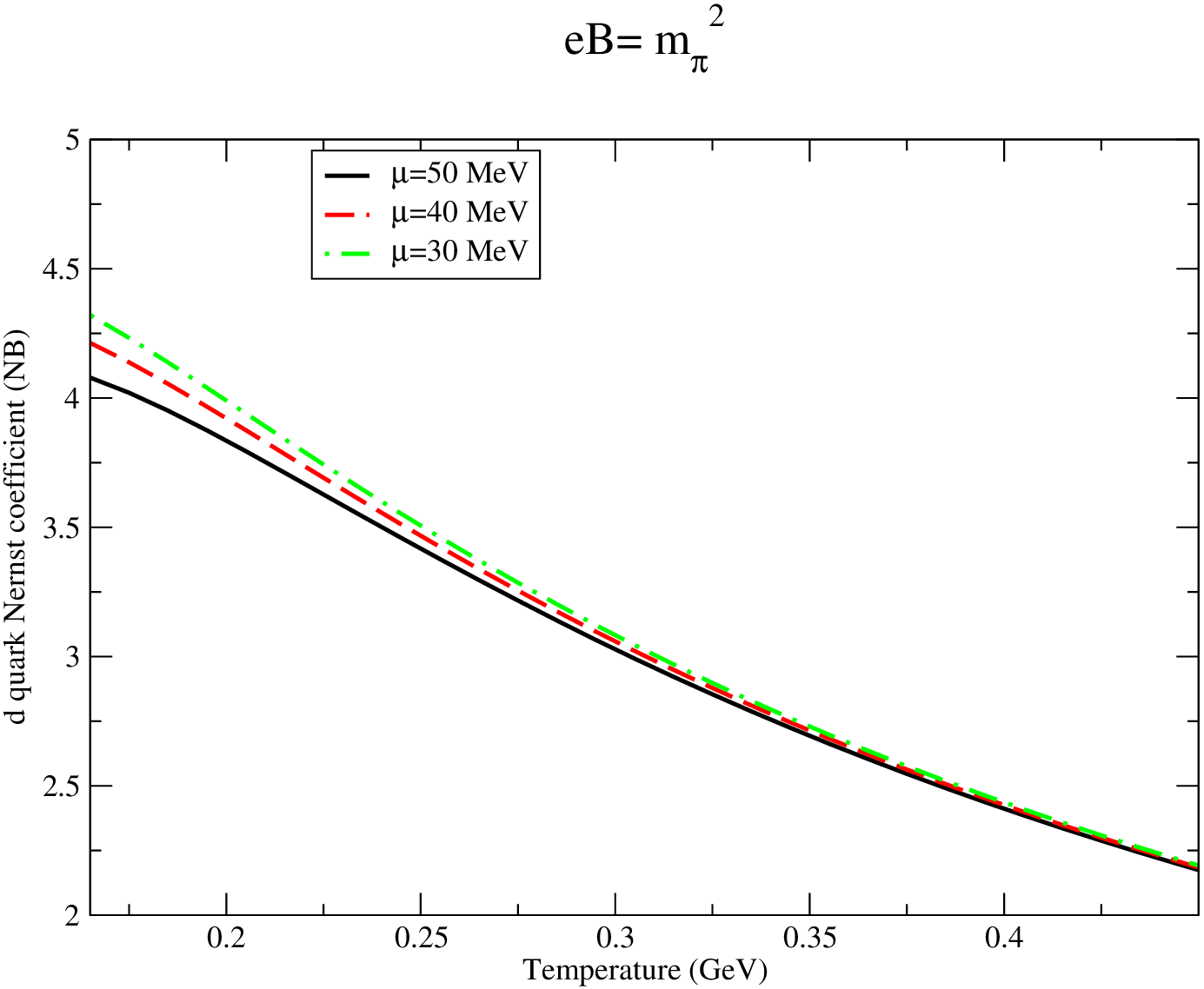}
		\caption{}\label{fig11b}
	\end{subfigure}
	\caption{Variation of Nernst coefficient of 
		$u$ (a) and $d$ (b) quarks with 
		temperature for different fixed values of quark chemical potential.}\label{fig11}
\end{figure}
Figures \eqref{fig11a} and \eqref{fig11b} show the individual Nernst coefficients for the medium composed exclusively of $u$ quarks and $d$ quarks respectively. The magnitude decreases almost monotonically with temperature for the entire temperature range. The impact of chemical potential is overall feeble, with it being discernible only near the transition temperature. From $2-2.5$ $T_c$, there is negligible impact of chemical potential on the value of the coefficients. The almost identical values of Nernst coefficient for both the $u$ quark medium and the $d$ quark medium suggests that the value is fairly independent of the quantum of charge carried by the individual charge carrier. This is unlike the case of Seebeck coefficient where the difference in the quantum of charge carried by the $u$ and $d$ quarks is directly reflected in their respective Seebeck coefficients.   
\begin{figure}[H]
	\begin{subfigure}{0.48\textwidth}
		\includegraphics[width=0.95\textwidth]{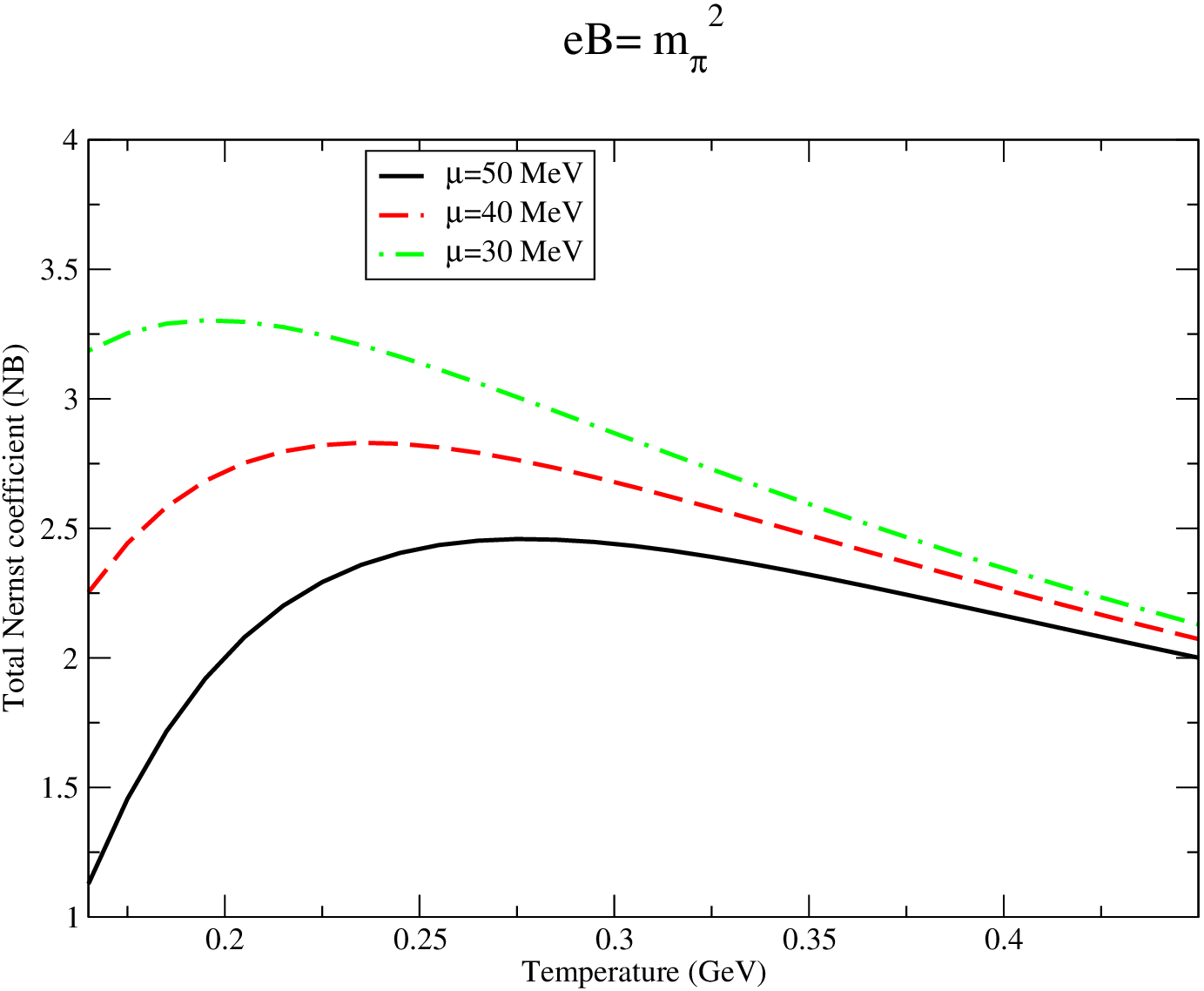}
		\caption{}\label{fig12a}
	\end{subfigure}
	\hspace*{\fill}
	\begin{subfigure}{0.48\textwidth}
		\includegraphics[width=0.95\textwidth]{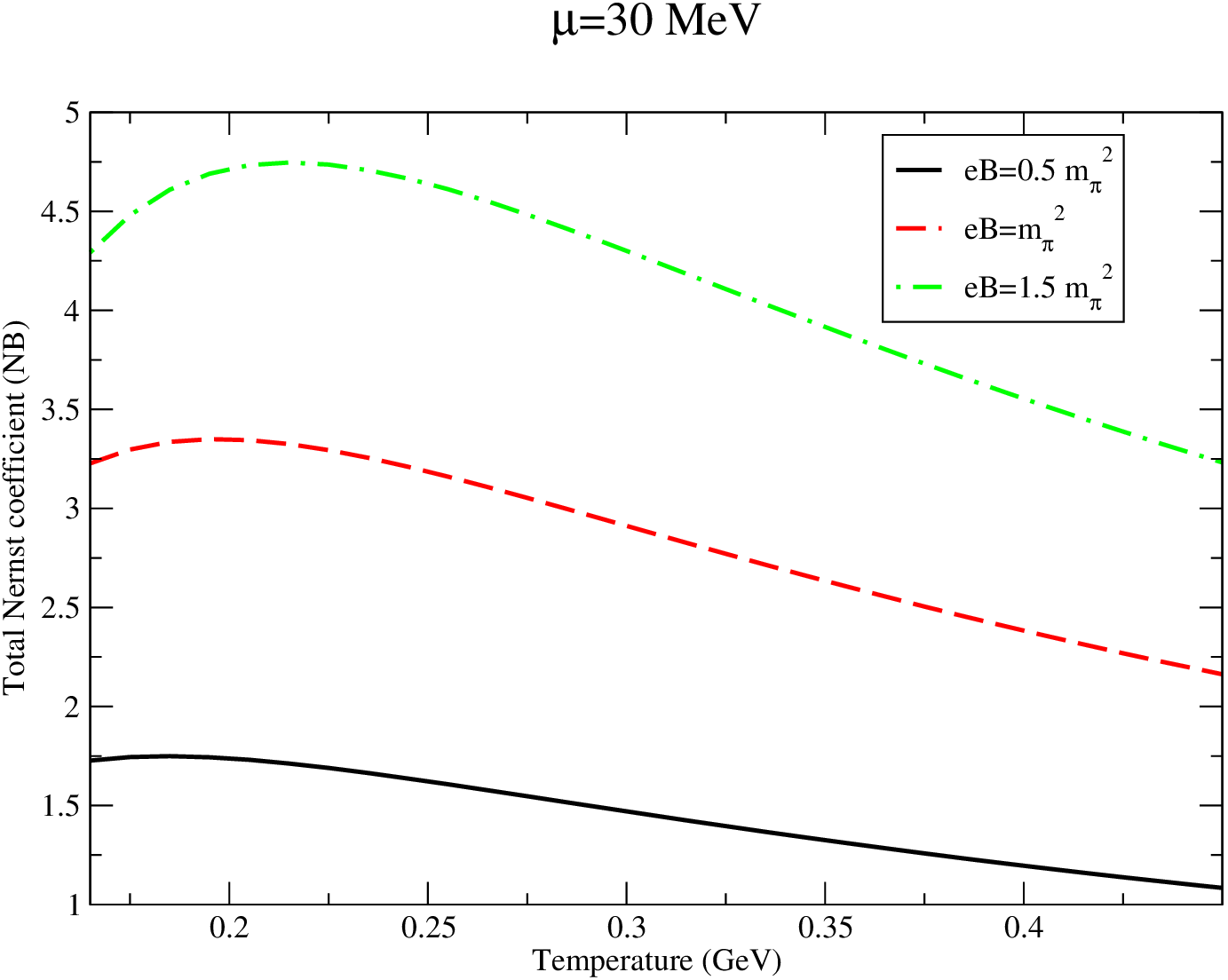}
		\caption{}\label{fig12b}
	\end{subfigure}
	\caption{\textbf{Left panel:} Variation of total Nernst coefficient with temperature at $eB=m_{\pi}^2$ for different fixed values of chemical potential. \textbf{Right panel:} Variation of total Nernst coefficient with temperature at $\mu=30$ MeV for different fixed values of magnetic field.}\label{fig12}
	\end{figure}

\begin{figure}[H]
	\centering
	\includegraphics[width=9.2cm,height=6.4cm,keepaspectratio]{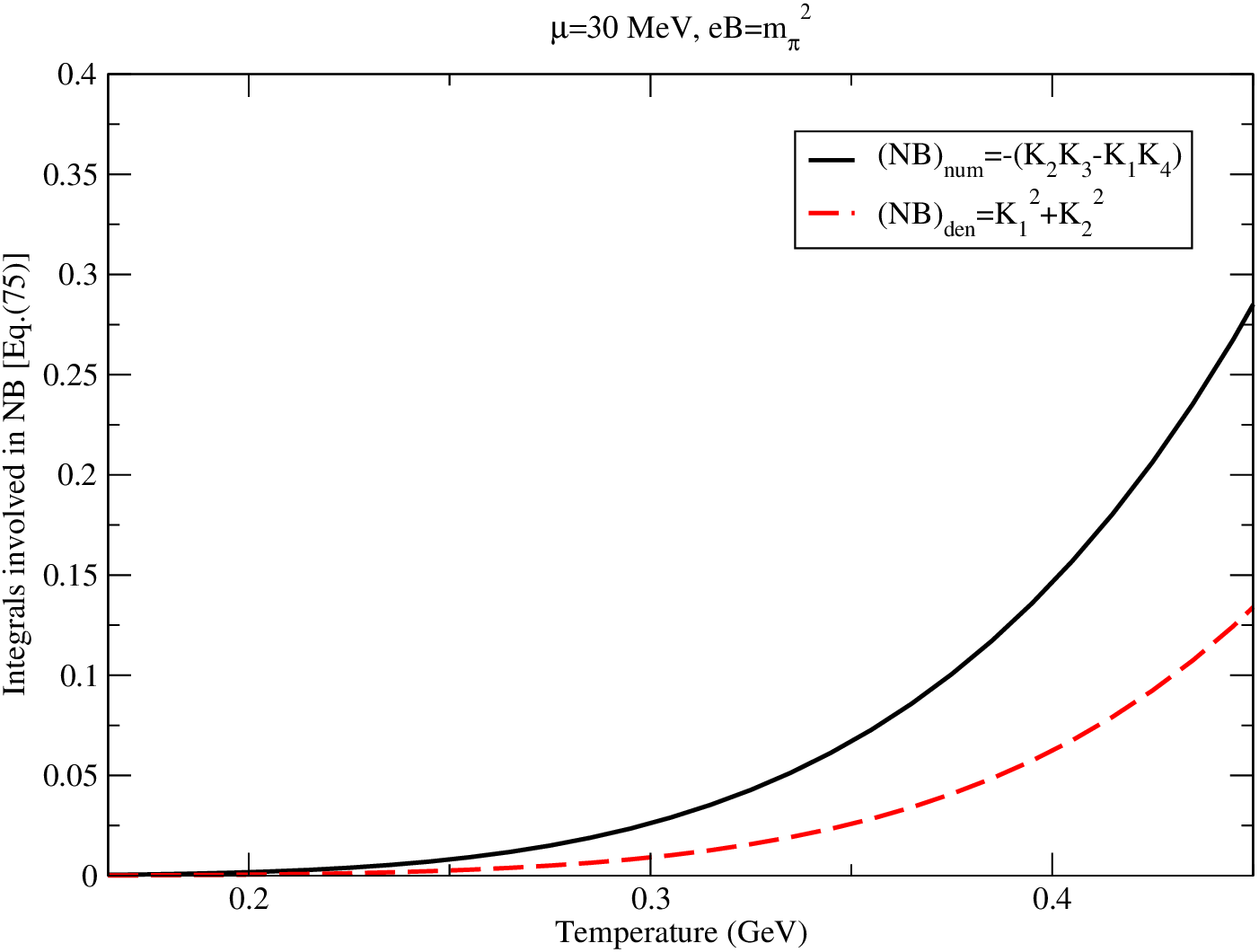}
	\caption{ Variation of numerator and denominator of total Nernst coefficient [Eq.\eqref{65}] with temperature for $eB=m_{\pi}^2$, $\mu=30$ MeV.}\label{fig13}
	\end{figure}

Fig.\eqref{fig12a} and \eqref{fig12b} show the variation of total Nernst coefficients with temperature for different values of chemical potential and magnetic fields respectively. As can be seen in Fig.\eqref{fig12a}, the coefficient records an increase with temperature near the crossover temperature after which the slope of the curve changes sign and the coefficient records a decreasing trend with temperature. Both the numerator and the denominator of the Nernst coefficient [Eq.\eqref{65}] are increasing functions of temperature for a fixed chemical potential. This is shown in Fig.\eqref{fig13}. The rate of increase of the denominator vis-a-vis the numerator is however greater at higher temperatures compared to lower temperatures. This leads to the ratio showing an increasing trend with temperature up to around $T=200$ MeV. Thereafter, the slope of the denominator increases faster with temperature, which leads to a reversal of the trend of the ratio with temperature [Fig.\eqref{fig12a}]. This reversal of slope of the total Nernst coefficient depends on the chemical potential. For a higher value of chemical potential, the temperature at which the slope of the coefficient reverses is also higher, as can be seen from Fig.\eqref{fig12a}. Fig.\eqref{fig12b} shows that the Nernst effect is more pronounced at higher values of magnetic field for a fixed chemical potential. This is understandable as a stronger magnetic field aids the drift of the charge carriers transverse to the temperature gradient. It should be noted that the individual as well as total Nernst coefficients come out to be zero for $B=0$, as expected. 

\noindent Seebeck and Nernst coefficients have been calculated for both the hadron gas medium and the quark-gluon plasma medium using a 2-dimensional formulation and relevant ansatzs.  Thermoelectric response in a hadron gas medium modelled by the Hadron Resonance Gas (HRG) model has been investigated in \cite{Das:PRD102'2020}, the obvious difference with our work therefore being the choice of the medium. They record a negative total Seebeck coefficient with an increase in the absolute value with temperature along with a decrease in the absolute values with increasing magnetic field. The behaviour is thus different (both the sign and the trend) from that obtained by us for the QGP medium. 
The variation of Nernst coefficient with temperature and chemical potential is roughly similar to the results obtained by us, with the difference being the initial increase in the coefficient magnitude (near the transition temperature) in our work. Seebeck coefficient (along with thermal and electrical conductivities) in a baryon asymmetric hot quark matter in the absence of magnetic field has been evaluated recently in \cite{Aman:arxiv}, wherein the authors report a negative Seebeck coefficient whose absolute value increases monotonically with temperature and decreases with chemical potential. Similar to our study, the medium considered by them consists of massive $u$ and $d$ quarks. However, they make use of the Nambu-Jona Lasinio (NJL) model for modelling the interactions and  calculation of the relaxation time of the medium, where, they consider quark-quark, quark-antiquark and antiquark-antiquark scattering processes mediated by $\sigma$ and $\pi$ meson exchanges. Further, they consider the spatial gradient of the quark chemical potential to be non-zero. This is in contrast to our work where scatterings mediated by thermally screened gluon exchange has been considered for the evaluation of the relaxation time, and the quark chemical potential is considered to be space independent.  These could be the reasons for the differences in the results obtained in the two studies apart from the consideration of a finite magnetic field in our study.  The above two comparisons show that the nature of the medium and the interactions considered therein affect both the direction of the induced electric field (sign of the Seebeck coefficient) as well as its behaviour with temperature and magnetic field. In \cite{kurian:PRD103'2021}, thermoelectric effects have been studien in the quark-gluon plasma medium possessing a finite magnetic field. The medium interactions have been incorporated using the Effective fugacity Quasiparticle model (EQPM) with zero current quark masses for the $u$ and $d$ quarks, whereas in our work, a finite current quark mass has been considered for both the quarks and interactions manifest via the quasiparticle masses obtained from perturbative QCD calculations. The other major difference is the ansatz [Eq.\eqref{49}] used for the calculation. The ansatz used by us is a natural generalization of the 1-D ansatz that has been used in multiple works before.  The magnitudes of the total Seebeck coefficient obtained in the temperature range is comparable to the ones obtained in our study. However, the sign of the coefficient is negative. This could be because of the difference in ansatz or modelling of in-medium interactions. The variation of Nernst coefficient with temperature and chemical potential bears  rough resemblances with our study except for the slight increase of $NB$ near $T_c$ reported by us. In \cite{Zhang:EPJC81'2021}, a similar study has been carried out, however, for an anisotropic QGP medium wherein the anisotropy of the medium has been incorporated via an anisotropy parameter $\xi$ in the distribution function. Further, another point of difference with our work is the expressions for total Seebeck and Nernst coefficients obtained from the individual ones. They have reported a negative total Seebeck coefficient with the magnitude decreasing with temperature and increasing with magnetic field. The Nernst coefficient reported by them bears a close resemblance with our study with its temperature variation recording a slight increase near $T_c$, similar to our work. This could be because of the similar structure of the ansatz used.

\section{Summary and Conclusions}
In this paper, we have investigated the thermoelectric phenomena of Seebeck effect and Nernst effect in a 
deconfined plasma of quarks and gluons in the presence of a magnetic field. Although large 
magnetic fields are produced in noncentral heavy ion collisions, only a small fraction of 
this magnetic field is expected to exist in a thermalised, strongly interacting QGP near 
$T_c$ due to the finite, small electrical conductivity of the created medium. By allowing for only 
small deviations of the system from equilibrium, we implicitly constrain the magnetic field 
to be not very strong. We have carried out the aforesaid analysis in the kinetic theory framework 
by applying the Boltzmann transport equation for a relativistic system in the relaxation 
time approximation wherein we assume that the phase space and dispersion relations of quarks are not affected by magnetic field via Landau quantization. We have quantified the thermoelectric response of the medium by calculating 
the Seebeck coefficient which is the induced electric field per unit temperature gradient 
of the medium. To this end, we have undertaken a two part approach wherein we first calculate only the Seebeck coefficient in a 1-D formulation and compare the results with that of the strong magnetic field case, also evaluated in a 1-D formulation in our previous work. Next, we adopt a 2-D approach and calculate the Seebeck and Nernst coefficients again, the 2-D formulation being necessitated by the study of the transverse Hall type effect, \emph{i.e.} the Nernst effect. 

\noindent From the 1-D study, our work shows that in the 
presence of a weak magnetic field, the mass of a particle has an amplifying effect on the individual 
Seebeck coefficients as can be seen from comparing Fig.\eqref{fig1b} and \eqref{fig2}, which might not 
be immediately obvious from Eq.\eqref{43}. We have commented on the temperature sensitivity of the individual Seebeck coefficients as a function of the background magnetic field strength. We have also tried to analyse the effect of particle mass on the temperature sensitivity of the individual coefficients in different domains of magnetic field strengths. We find that the physical medium consisting of different species of 
quarks, in the presence of a weak magnetic field has a feeble thermoelectric response in the entire temperature range considered, much less than that in the case of a strong magnetic field, as can be seen in Fig.\eqref{fig7}. The weighted average prescription of calculating the total Seebeck coefficient can lead to interesting results for the total coefficient. As can be seen in Fig.\eqref{fig7}, the direction of induced electric field is opposite to the direction of temperature gradient in the presence of a strong magnetic field while for the weak $B$ case, the induced electric field and the temperature gradient are in the same direction. We have observed that in the presence of weak magnetic field, the individual seebeck coefficient magnitude gets amplified with increasing particle mass, unlike in the case of strong magnetic field where we see a suppression. We have also found that the temperature sensitivity of the individual coefficient increases with increase in the particle mass in the presence of a weak magnetic field, whereas the same decreases in the presence of a strong magnetic field.

\noindent From our 2-D results of the Seebeck coefficient, it is clear that both the individual and the total Seebeck coefficients follow the same trends with temperature, chemical potential and magnetic field as their 1-D counterpart. The major difference is that in the 2-D formulation it is no longer possible to write the total Seebeck coefficient of the medium as a simple weighted average of the individual Seebeck coefficients. For the Nernst coefficients ($NB$), we find that the sign of the individual Nernst coefficient is independent of the charge of the majority charge carrier of the medium, unlike in the case of the individual Seebeck coefficients. Also unlike the Seebeck coefficient, the Nernst coefficient decreases with chemical potential. The total Nernst coefficient records a slight increase in magnitude with temperature near $T_c$. The slope of the curve then becomes negative and thus the magnitude decreases with temperature thereafter. The temperature at which this reversal of slope takes place is higher for a higher value of chemical potential.  

\section{Acknowledgement}
B. K. P. is thankful to the Council of Scientific and Industrial Research (Grant No.
03(1407)/17/EMR-II) for the financial assistance. D.D. would like to thank Shubhalaxmi Rath for valuable discussions regarding various aspects of the paper.

\end{document}